\documentclass[fleqn]{2023SCGE}
\usepackage[dvipsnames]{xcolor} 
\usepackage{orcidlink} 
\usepackage{ulem}
\begin{document}

\ensubject{subject}

\ArticleType{Article}
\SpecialTopic{SPECIAL TOPIC: }
\Year{2025}
\Month{June}
\Vol{X}
\No{1}
\DOI{??}
\ArtNo{000000}
\ReceiveDate{June 11, 2025}
\AcceptDate{xxx x, 2025}
\OnlineDate{xxx x, 2025}
\renewcommand\floatpagefraction{.9}
\renewcommand\topfraction{.9}
\renewcommand\bottomfraction{.9}
\renewcommand\textfraction{.1}
\setcounter{totalnumber}{50}
\setcounter{topnumber}{50}
\setcounter{bottomnumber}{50}

\newcommand{\be}{\begin{equation}}
\newcommand{\ee}{\end{equation}}
\newcommand{\ba}{\begin{eqnarray}}
\newcommand{\ea}{\end{eqnarray}}
\newcommand{\no}{\nonumber}
\newcommand{\bi}{\begin{itemize}}
\newcommand{\ei}{\end{itemize}}
\newcommand{\kpch}{h^{-1} {\rm kpc}}
\newcommand{\mpch}{h^{-1} {\rm Mpc}}
\newcommand{\mpcht}{h^{-3} {\rm Mpc^3}}
\newcommand{\gpch}{h^{-1} {\rm Gpc}}
\newcommand{\gpcht}{h^{-3} {\rm Gpc^3}}
\newcommand{\hmpc}{h {\rm Mpc}^{-1}}
\newcommand{\hmpct}{h^3 {\rm Mpc}^{-3}}
\newcommand{\hgpc}{h{\rm Gpc}^{-1} }
\newcommand{\hgpct}{h^3 {\rm Gpc}^{-3} }
\newcommand{\rmnum}[1]{\romannumeral #1}
\newcommand{\Rmnum}[1]{\uppercase\expandafter{\romannumeral #1}}

\newcommand{\Msun}{M_{\odot}}
\newcommand{\Kun}{\textsc{Kun} }
\newcommand{\diff}{\mathrm{d}}
\newcommand{\rockstar}{\textsc{Rockstar} }
\newcommand{\fof}{\textsc{FoF} }
\newcommand{\subfind}{\textsc{SubFind} }

\title{CSST Cosmological Emulator II: Generalized Accurate Halo Mass Function Emulation}{CSST Cosmological Emulator II}

\author[1,2,3,4]{Zhao Chen\orcidlink{0000-0002-2183-9863}}{} 
\author[2,3,4]{Yu Yu\orcidlink{0000-0002-9359-7170}}{{}}

\thanks{Corresponding author (email:
~\href{chyiru@sjtu.edu.cn}{chyiru@sjtu.edu.cn};
~\href{yuyu22@sjtu.edu.cn}{yuyu22@sjtu.edu.cn})}

\AuthorMark{Chen Zhao}

\AuthorCitation{Chen Z, Yu Y, et al.}

\address[1]{Tsung-Dao Lee Institute, Shanghai Jiao Tong University, Shanghai 200240, China}
\address[2]{Department of Astronomy, School of Physics and Astronomy, Shanghai Jiao Tong University, Shanghai 200240, China}
\address[3]{State Key Laboratory of Dark Matter Physics, School of Physics and Astronomy, Shanghai Jiao Tong University, Shanghai 200240, China}
\address[4]{Key Laboratory for Particle Astrophysics and Cosmology (MOE)/Shanghai Key Laboratory for Particle Physics and Cosmology, Shanghai 200240, China}


\abstract{
Accurate theoretical prediction for halo mass function across a broad cosmological space is crucial for the forthcoming Chinese Space Station Survey Telescope (CSST) observations, which will capture cosmological information from multiple probes, e.g., cluster abundance, and weak lensing.
In this work, we quantify the percent-level impact of different mass binning schemes when measuring the differential halo mass function from simulations, and demonstrate that the cumulative form of the halo mass function is independent of the binning scheme.
Through the recently finished \Kun simulation suite, we propose a generalized framework to construct multiple accurate halo mass function emulators for different halo mass definitions, including $M_{200m}$, $M_{vir}$, and $M_{200c}$.
This extends our \texttt{CSST Emulator} to provide fast and accurate halo mass function predictions for halo mass $M\geq 10^{12}\,h^{-1}\Msun$ up to $z=3.0$.
For redshifts $z\leq 1.0$, the accuracy is within $2\%$ for $M\leq 10^{13}\,h^{-1}\Msun$, $5\%$ for $M\leq 10^{14}\,h^{-1}\Msun$, and $10\%$ for $M\leq 10^{15}\,h^{-1}\Msun$, which is comparable with the statistical errors of training simulations.
This tool is integrated in \texttt{CSST Emulator} and publicly available at \url{https://github.com/czymh/csstemu}, providing a fast and accurate theoretical tool to obtain unbiased cosmological constraints of the upcoming CSST survey.
}



\maketitle


\begin{multicols}{2}

\section{Introduction}
\label{sec:intro}

Dark matter halos are the fundamental unit of the cold dark matter (CDM) universe, which are formed through three-dimensional gravitational collapse.
Their abundance, properties, distribution, and nonlinear evolution are crucial for the modern cosmological analysis, providing constraints on the evolution of dark energy and dark matter, the impact of massive neutrinos, and even the galaxy-halo connection (e.g.,~\cite{2001ApJ...561...13B,2003A&A...398..867S,2009ApJ...692.1060V,2017MNRAS.470..551Z,2018A&A...620A...1M,2019MNRAS.482.1352M,2019ApJ...878...55B,2024A&A...682A.148F}).
Especially, massive dark matter halos are commonly connected with galaxy clusters, the largest bounded structures in the Universe.
The observed number counts of clusters, identified by various surveys and quantified through halo mass function (HMF), serves as a powerful bridge between cosmological theory and observations (see reviews by~\cite{2011ARA&A..49..409A,2012ARA&A..50..353K,2013PhR...530...87W,2025arXiv250507697M}).
For the past two decades, cluster samples have been detected from the optically imaging data (e.g., SDSS~\cite{2014ApJ...785..104R}, DES~\cite{2022PhRvD.105b3520A}, KiDS~\cite{2022A&A...659A..88L}),
the X-ray observations (e.g., ROSITA PSPC survey~\cite{2007ApJS..172..561B}, XXL survey~\cite{2016A&A...592A...2P}, eROSITA~\cite{2022A&A...661A...7B}),
and the Sunyaev-Zel'dovich effect catalogs from the cosmic background radiation (CMB) surveys (e.g., Planck~\cite{2016A&A...594A..27P}, ACT~\cite{2018ApJS..235...20H}, SPT~\cite{2020AJ....159..110H}).
These observations have enabled cluster analysis to become an informational cosmological probe for understanding the structure growth history of the Universe (e.g., \cite{2015MNRAS.446.2205M,2019MNRAS.489..401Z,2021PhRvD.103d3522C}).
In the near future, the Stage-IV surveys, e.g, the Vera Rubin Observatory Legacy Survey of Space and Time (LSST\footnote{\url{http://www.lsst.org}} \cite{2009arXiv0912.0201L}), the Euclid satellite\footnote{\url{http://www.euclid-ec.org}}~\cite{2011arXiv1110.3193L,2024arXiv240513491E}, the Nancy Grace Roman Space Telescope (Roman\footnote{\url{https://roman.gsfc.nasa.gov/}}~\cite{2019BAAS...51c.341D}), and the Chinese Space Station Survey Telescope (CSST\footnote{\url{https://www.nao.cas.cn/csst/}}~\cite{2019ApJ...883..203G}), will obtain large samples of optical clusters to tighten the cosmological constraints on $\Omega_m$ and $\sigma_8$.
The combination of spectroscopic and photometric survey of CSST could detect $200,000\sim 400,000$ clusters in $17,500 \deg^{2}$ survey area at $z\in[0,\,1.5]$ and provide precise estimations of cluster redshift and mass~\cite{2023MNRAS.519.1132M,2023RAA....23d5011Z,2025SCPMA..6880402G,2025arXiv250704618C}.
The Jiao Tong University Spectroscopic Telescope (JUST~\cite{2024AstTI...1...16J}) cluster cosmology survey aims to provide a complete spectroscopic cluster sample ($\sim50,000$) for galaxies at $r<20$ mag and $z<0.6$.
The unprecedented observational precision demands percent-level accuracy of HMF predictions to gain unbiased and accurate constraints on cosmological parameters.

Early HMF are derived analytically through assuming the spherical collapse in a linear matter field (e.g., Press-Schechter~\cite{1974ApJ...187..425P} and excursion set theory~\cite{1991ApJ...379..440B}).
In these predictions, the cosmological dependence of HMF is expected to be captured mostly by the underlying matter density fluctuations $\sigma(M,z)$, indicating the universality of the HMF.
Over the past years, this universality is validated through various $N$-body simulations and fitted to provide predictions of halo numbers (e.g.,~\cite{2001MNRAS.321..372J,2003MNRAS.346..565R,2010MNRAS.403.1353C,2012MNRAS.426.2046A,2013MNRAS.433.1230W,2016MNRAS.456.2361B}).
However, many other works have detected $10\sim20\%$ deviations from the universality, which is dependent on the redshifts, halo mass definitions, and algorithms of halo finding (e.g.,~\cite{2008ApJ...688..709T,2016MNRAS.456.2486D,2020ApJ...903...87D,2022MNRAS.509.6077O})
The non-universality of mass functions on redshift, cosmology, and the ambiguity of halo catalogs can potentially become the dominant systematic errors for future galaxy cluster studies (e.g.,~\cite{2020A&A...643A..20S,2021A&A...649A..47A,2023A&A...671A.100E}).
Moreover, this influence may increase to a higher level for the cosmologies with non-cosmological constant dark energy (e.g.,~\cite{2011ApJ...732..122B}), which catches much attention due to recent DESI constraints (e.g.,~\cite{2025JCAP...02..021A,2025arXiv250314738D}).

Fortunately, the non-universality across wide ranges of cosmological parameters and redshifts can be solved by constructing emulators.
The cosmological emulator for a specific summary statistic is trained to provide accurate high-dimensional interpolations based on a finite number of expensive simulations whose cosmologies are exquisitely designed in a given parameter space.
This is achieved by learning the connection between the input cosmological parameters and the output summary statistic, which is commonly done through the Gaussian Process Regression (GPR) method and other machine learning techniques (see short reviews in~\cite{2022LRCA....8....1A,2023RPPh...86g6901M}).
Serval emulators have been constructed for HMF (e.g., Aemulus~\cite{2019ApJ...872...53M}, DarkQuest~\cite{2019ApJ...884...29N}, Mira-Titan~\cite{2020ApJ...901....5B} and Aemulus-$\nu$~\cite{2025JCAP...03..056S}), and other statistics, 
such as the nonlinear matter power spectrum (e.g.,~\cite{2010ApJ...715..104H,2014ApJ...780..111H,2019MNRAS.484.5509E,2021MNRAS.505.2840E,2023MNRAS.520.3443M,Chen2025}), 
linear halo bias (e.g.,~\cite{2019arXiv190713167M,2019ApJ...884...29N}), 
galaxy clustering (e.g.,~\cite{2015ApJ...810...35K,2019ApJ...874...95Z,2020MNRAS.492.2872W,2022MNRAS.515..871Y,2023ApJ...948...99Z,2024ApJ...961..208S}), 
basis spectra of bias expansion model (e.g.,~\cite{2021JCAP...09..020H,2023MNRAS.524.2407Z,2023MNRAS.520.3725P,2023JCAP...07..054D,2025arXiv250604671Z}), 
and a variety of high-order statistics (e.g.,~\cite{2018JCAP...03..049L,2019PhRvD..99h3508L,2019PhRvD..99f3527L,2019JCAP...05..043C,2025PhRvD.111d1302W}).

The halo mass corresponds to the total mass enclosed within a sphere where the mean density exceeds a specified density threshold $\Delta$.
This definition encompasses several conventional variants:
$M_{vir}$ adopts a redshift-dependent virial overdensity $\Delta_{vir}$ to demarcate the fully collapsed regions (e.g.,~\cite{1998ApJ...495...80B}).
$M_{200c}$ and $M_{200m}$, widely used in theoretical and observational studies(e.g.,~\cite{2016MNRAS.456.2361B,2019ApJ...878...55B}), employ thresholds of 200 times the critical density $\rho_c$ and 200 times the mean matter density ($\rho_{cb}$, excluding neutrinos), respectively.
These distinct mass definitions yield systematically divergent halo properties and fundamentally influence cosmological interpretations.
While most previous HMF emulators only predict mass function for a single halo definition given by a specific halo finder (e.g., $M_{200m}$~\cite{2019ApJ...872...53M,2019ApJ...884...29N,2025JCAP...03..056S}, $M_{200c}$~\cite{2020ApJ...901....5B}). 
In this article, we propose a generalized framework to construct multiple accurate halo mass function emulators for different halo mass definitions, including $M_{200m}$, $M_{vir}$, and $M_{200c}$.
The accuracy of emulators for different halo definitions is consistent, indicating the same level of systematic uncertainties independent of halo ambiguity.
The HMF emulator here is a straightforward extension of \texttt{CSST Emulator}~\cite{Chen2025}, based on the \Kun simulation suite across a broad cosmological parameter space.
The final emulator can predict HMF for halo mass $M\geq 10^{12}\, h^{-1}\Msun$ up to $z\leq 3.0$.
The overall uncertainties lie within the statistical errors of training simulations.

The rest of the article is structured as follows: 
Section~\ref{sec:hmf} provides a brief review of the theoretical background of the HMF.
Especially, section~\ref{sec:binning} presents the impact of the binning effect on the differential HMF and proves that the cumulative HMF is a much better choice for emulation.
We then describe the \Kun simulation suite and quantify the influence of mass resolution on the HMF measurements.
Section~\ref{sec:emulation} demonstrates the detailed framework of the emulator construction and validates the accuracy of the emulator through Leave-One-Out errors.
Section~\ref{sec:validation} validates the excellent performance of \texttt{CSST Emulator} compared with other emulators and fitting formulae.
Finally, in Section~\ref{sec:conc}, we conclude this work and discuss the future directions.

\section{Halo Mass Function}
\label{sec:hmf}

HMF is a fundamental quantity in cosmology, which is widely used to characterize the abundance of dark matter halos in the Universe.
In this section, we briefly review the theoretical background of the HMF and discuss the binning effect, a potential systematic when measuring HMF from simulations or observations.

\subsection{Theoretical background}
\label{sec:hmf-theory}

The differential HMF is defined as the number density of halos in given (logarithmic) mass bins.
\begin{equation}
\label{eq:hmf}
\frac{\diff n}{\diff \ln M} \diff \ln M = \frac{\rho_{\mathrm{cb}}}{M} \nu f(\nu) \diff \ln \nu\ .
\end{equation}
Here, $\rho_\mathrm{cb}$ is the mean matter density excluding neutrino mass, $M$ is the halo mass, and $\nu f(\nu)$ is the multiplicity function.
Massive neutrinos barely cluster on halo scales and thus contribute little to the gravitational collapse of dark matter halos.
Therefore, we use cold dark matter plus baryon density (`cb') in the prescription of HMF~\cite{2014JCAP...02..049C}.
Many previous works discussed the universality and non-universality of HMF based on this multiplicity function.
The peak height $\nu = \delta_c(z) / \sigma(M,z)$, where the $\delta_c(z)$ is the time-evolving critical overdensity threshold for spherical collapse in the linear theory.
In this work, we take the expression in Kitayama \& Suto 1996~\cite{1996ApJ...469..480K},
\begin{equation}
\label{eq:delta_c}
\delta_{\mathrm{c}}=\frac{3}{20}(12 \pi)^{2 / 3}\left[1+0.0123 \log _{10} \Omega_{\mathrm{cb}}(z)\right]\ ,
\end{equation}
where $\Omega_{\mathrm{cb}}(z)$ is cold dark matter plus baryons density at redshift $z$.
$\sigma(M,z)$ is the mass variance of the linear `cb' density field with a spherical top-hat smoothing, and can be derived from the linear `cb' power spectrum $P_{\mathrm{cb}}(k,z)$ through
\begin{equation}
\label{eq:sigma}
\sigma^2(M, z)=\frac{1}{2 \pi^2} \int_0^{\infty} \diff k k^2 P_{\mathrm{cb}}(k, z) W^2(k R)\ .
\end{equation}
Here, $R(M) = (3M/4\pi\rho_{m})^{1/3}$ represents the Lagrangian radius for a given halo mass $M$, and $W(k R)$ is the 3D spherical top-hat window function in the Fourier space,
\begin{equation}
\label{eq:WkR}
W(x) = 3(\sin x - x \cos x) / x^3\quad \text{with} \quad x = k R\ .
\end{equation}

Many previous studies have demonstrated that the multiplicity function tends to be universal across different cosmologies.
This indicates that the cosmological dependence of the HMF can be effectively captured by the matter power spectrum or the peak height.
Based on this, many fitting formulas have been proposed to provide accurate predictions of the HMF.
However, the non-universality has been observed by recent works in a broader cosmological space, whose amplitude is related to the redshift and halo definition.
In this work, we aim to provide a percent-level accurate emulator of HMF for different halo mass definitions up to $z=3.0$ in a wide cosmological parameter space.
First of all, we need accurate HMF measurements from $N$-body simulations.
Unfortunately, the measured differential HMF is dependent on the specific binning scheme.
In the next section, we describe the reason and quantify the amplitude of the binning effect.

\subsection{Binning effect}
\label{sec:binning}

HMF is typically measured by counting the number of halos in given mass bins in logarithmic space.
The measured result is different when using different bin widths.
Courtin et al. 2011~\cite{2011MNRAS.410.1911C} found that the measured HMF is less sensitive to the bin width when the bin-center mass is used instead of the averaged halo mass over the bin.
However, the choice of bin width can still cause percent-level bias for the HMF measurement.
It is crucial to quantify the amplitude of this binning effect as the emulator is aiming at a similar accuracy level.

\begin{figure}[H]
    \centering
    \includegraphics[width=0.4\textwidth]{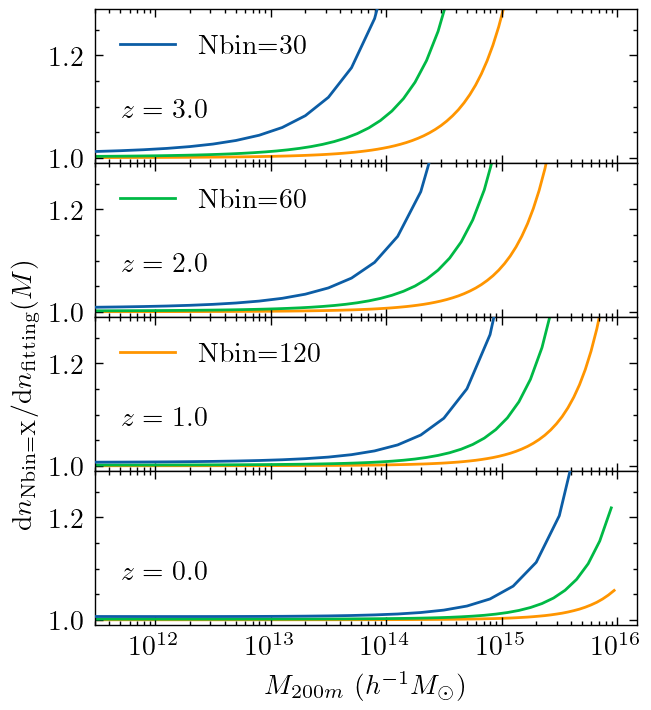}
    \caption{The comparison between the binned and `true' HMF for three different binning schemes.
    From bottom to top panel, the impact of the binning effect increases with the increased redshift.       
    }
    \label{fig:fit_binning}
\end{figure}

The origin of the binning effect is the uncertainty of the `true' halo mass in each bin.
Li \& Smith 2024~\cite{2024arXiv241118722L} quantified this effect on the multiplicity function and observed percent-level implications in their Appendix A.
They suggest the binning effect should be accounted for when using fewer than 10 bins per decade in the measurement.
However, an increasing number of bins leads to increased statistical errors and potentially decreased emulator accuracy, especially for larger halos.
In this work, we use a similar method to quantify the impact on the differential HMF measurement.
Firstly, we assume the distribution of halo mass follows the fitting formula in Tinker et al. 2008~\cite{2008ApJ...688..709T} (hereafter Tinker08, also T08 in equations).
The halo mass is defined as $200$ times the mean matter density, $M_{200m}$.
The `true' value of halo mass function at a specific 
mass $M_{i}$
is given by $\frac{\diff n_{\rm T08}}{\diff \ln M}|_{M=M_{i}}$.
Then, we calculate the binned HMF by integrating the fitting formula over the mass bin, $\frac{1}{\Delta\ln M}\int_{\ln M_{i}-\Delta\ln M/2}^{\ln  M_{i}+\Delta\ln M/2} \frac{\diff n_{\rm T08}}{\diff \ln M}(M) \diff \ln M $.
The relative difference between the binned and `true' HMF is shown in Fig.~\ref{fig:fit_binning}.
We chose three different binning schemes in the same mass range $[10^{10}, 10^{16}]\, h^{-1}\Msun$.
$\mathrm{Nbin}=60$ with bin width $0.1$ dex is our fiducial choice, also used in the HMF measurements of \Kun simulations in Section~\ref{sec:emulation}.
The other two binning schemes are $\mathrm{Nbin}=30$ with bin width $0.2$ dex, and $\mathrm{Nbin}=120$ with bin width $0.05$ dex.
We find that the relative difference reaches 5\% at $M_{200m}\sim 3\times10^{15}\, h^{-1}\Msun$ and $z=0.0$ for our fiducial case.
The bias increases as the redshift and halo mass increase.
As HMF has a slope inside the bin, the binned HMF measurement can be treated as the HMF on an averaged mass, which usually differs from the bin-center mass.
The steeper the HMF curves, the larger the difference between averaged halo mass and bin-center mass, hence leading to a larger systematic deviation.
For a sparser binning scheme, this impact becomes more significant.
This impact can be reduced by using a finer binning scheme, but the measurements will suffer larger sample variance, especially in the high-mass end.
The binning strategy is a trade-off between measurement fidelity and Poisson-limited uncertainties~\cite{2024MNRAS.535.3500Y}.

\begin{figure}[H]
    \centering
    \includegraphics[width=0.4\textwidth]{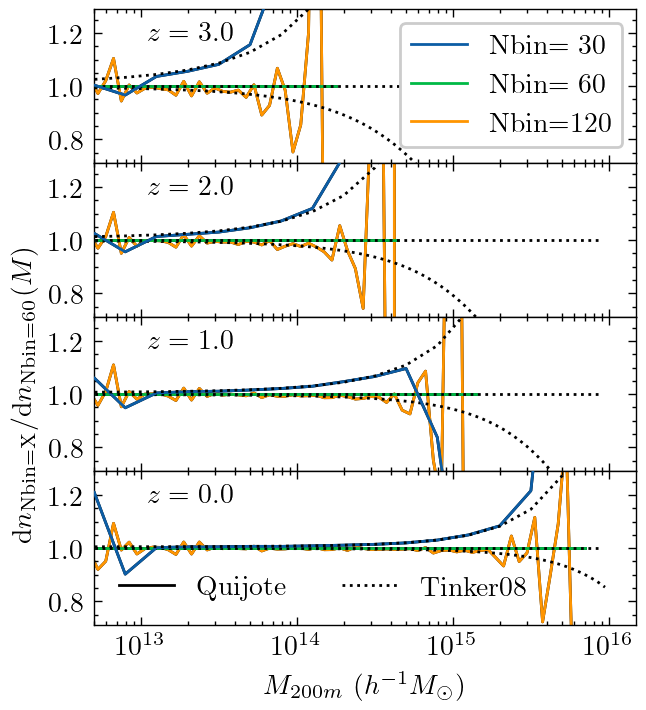}
    \caption{The differences between the binned HMF with different binning schemes measured from 100 Quijote-HR simulations (solid lines).
    Black dotted lines illustrate the results of the predicted bias using the Tinker08 fitting formula.
    }
    \label{fig:quijote_binning}
\end{figure}

To illustrate the robustness of this test, we use the same binning scheme to measure the HMF from simulations and compare it with the predicted bias above.
The averaged measurements of 100 fiducial\_HR simulations in the Quijote suite\footnote{\url{https://quijote-simulations.readthedocs.io/}}~\cite{2020ApJS..250....2V} (hereafter Quijote-HR) are utilized to reduce the sample variance.
Each simulation evolves $1024^3$ particles in a $1\,\gpcht$ cubic box with a different initial seed.
Then we compare the binned HMF in the cases of bin width $0.2$ dex and $0.05$ dex, with the fiducial case.
The results of averaged Quijote-HR measurements and theoretical predictions are represented by solid lines and dotted lines, respectively, in Fig.~\ref{fig:quijote_binning}.
The fluctuations for simulation results at large and small halo masses are caused by the dominant sample variance and the limited mass resolution, respectively.
We observe an excellent agreement between the measured bias in simulations and the predicted bias using Tinker08 fitting formula for all redshifts and halo masses.
On one hand, this indicates that the binning effect can be effectively mitigated by dividing the binned HMF and `true' HMF obtained from a more sophisticated fitting formula.
On the other hand, this result confirms that the binning effect originates from the uncertainties of variables, i.e., the halo mass in each bin.
To overcome this problem, a natural way is to use the cumulative halo mass function, of which the variables are fixed mass thresholds.

\subsection{Cumulative Halo Mass Function}
\label{sec:cumhmf}

\begin{figure}[H]
    \centering
    \includegraphics[width=0.4\textwidth]{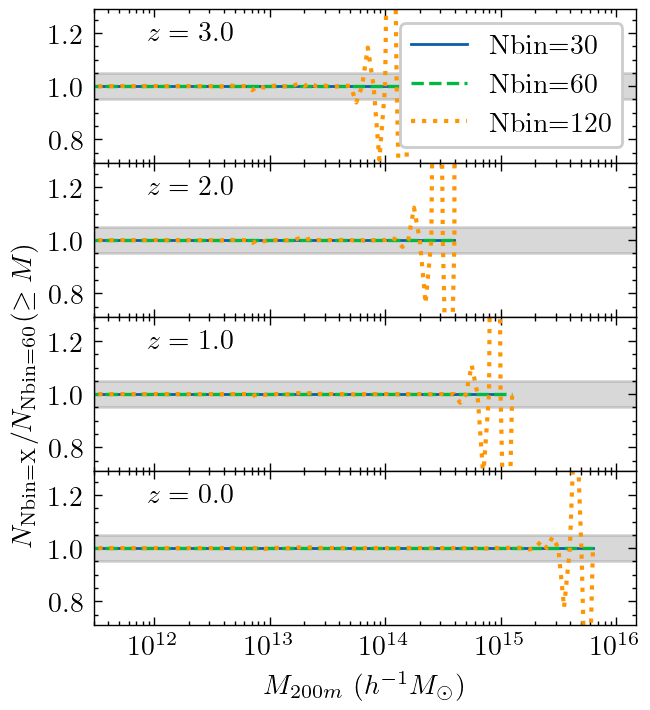}
    \caption{The comparison of the cumulative HMF for the fiducial case and the two other binning schemes measured from 100 Quijote-HR simulations at different redshifts.
    The gray region indicates $\pm 5\%$ difference.}
    \label{fig:N_binning}
\end{figure}

The cumulative halo mass function is defined as
\begin{equation}
\label{eq:cumhmf}
N(\geq M) = \int_{M}^{\infty} \frac{\diff n}{\diff \ln M} \diff \ln M \,.
\end{equation}
The variable of this form is a halo mass threshold instead of a mass bin.
Therefore, this quantity suffers no effect from the distribution of masses in the bins.
To confirm this, in Fig.~\ref{fig:N_binning} we illustrate the comparison of the cumulative HMF for the fiducial case and the two other binning schemes measured from 100 Quijote-HR simulations at different redshifts.
All the relative differences are invisible, except for the sample variance induced fluctuations at the highest mass end when comparing the finest binning scheme to the fiducial one (orange lines).
As a conclusion, the binning effect can introduce a significant influence on the measurements of differential HMF, especially at the high-mass end and high redshift.
Although a finer binning scheme can mitigate this impact, the increased sample variance will affect the final accuracy of the emulator, especially at the cluster mass scales.
Compared with the differential HMF, the cumulative one is smoother and suffers fewer statistical errors except for the highest mass bin.
Moreover, the slowly changing curves are also more friendly to be emulated.
In this work, we decide to use the cumulative HMF as the observable of interest.
The fiducial bin width in the following measurements is fixed at $0.1$ dex per decade, i.e., $\mathrm{Nbin}=60$.

\section{The \Kun simulation suite}
\label{sec:kun}

\subsection{Data Overview}
\label{sec:overview}

The \Kun simulation suite\footnote{\url{https://kunsimulation.readthedocs.io/}}~\cite{Chen2025} is a part of the \textsc{Jiutian} simulations~\cite{Han2025}, which is designed for the extragalactic surveys of CSST.
This suite consists of 129 high-resolution simulations with box size $L=1\,\gpch$ and evolving $3072^3$ particles.
Their cosmological parameters span a broad parameter space under the eight-dimensional $w_0 w_a \mathrm{CDM} + \sum m_{\nu}$ model.
The detailed cosmological parameters are summarized as follows.
\begin{equation}
    \label{eq:parameter-space}
    \begin{aligned}
    \Omega_{\mathrm{b}} & \in[0.04, 0.06]\ , \\
    \Omega_{\mathrm{cb}} & \in[0.24, 0.40]\ , \\
    n_{\mathrm{s}} & \in[0.92, 1.00]\ , \\
    A_{s} & \in[1.70, 2.50] \times 10^{-9}\ , \\
    H_{0} & \in[60, 80]\ \mathrm{km\, s^{-1}\, Mpc^{-1}}\ , \\
    w_{0} & \in[-1.30, -0.70]\ , \\
    w_{a} & \in[-0.50, 0.50]\ , \\
    \sum m_{\nu} & \in [0.00, 0.30]\ \mathrm{eV}\ .
    \end{aligned}
\end{equation}
Here, $\Omega_\mathrm{b}$ and $\Omega_\mathrm{cb}$ represent the baryon and cold dark matter plus baryon density at present.
$n_{s}$ and $A_{s}$ describe the spectral index and amplitude of the primordial power spectrum, respectively.
The expansion rate of the Universe at the current time is given by the Hubble parameter $H_{0}$.
And we use the CPL parametrization~\cite{2001IJMPD..10..213C,2003PhRvL..90i1301L} $w(a) = w_{0} + w_{a}(1 - a)$ to capture the dynamical evolution of the dark energy.
The impact of massive neutrinos on the cold dark matter is incorporated by the recent Newtonian motion gauge method~\cite{2020JCAP...09..018P,2022JCAP...09..068H}.
Only one single massive neutrino species is considered in this suite, and the total mass is given by $\sum m_{\nu}$.
We take the Planck 2018 cosmology as the fiducial cosmology (c0000) and utilize the Sobol sampling method~\cite{sobol1967distribution} to generate 128 training points (c0001-c0128).
In our first article~\cite{Chen2025}, this method exhibits an excellent performance in emulating the matter power spectrum, compared with the widely used Latin hypercube sampling method~\cite{mckayComparisonThreeMethods1979,tangOrthogonalArrayBasedLatin1993}.

All simulations are run by the modified Gadget-4\footnote{\url{https://wwwmpa.mpa-garching.mpg.de/gadget4/}} $N$-body solver with the fixed amplitude method~\cite{2016MNRAS.462L...1A} to suppress the cosmic variance at large scales.
The initial condition is generated by the second-order Lagrangian Perturbation Theory (2LPT) at the fixed redshift $z_{\rm ini}=127$.
The cluster cosmology mainly concerns halos with masses $\geq 10^{13}\,h^{-1}\Msun $.
Our mass resolution is $2.87 \frac{\Omega_\mathrm{cb}}{0.3} \times 10^{9}\, h^{-1}\Msun$, implying that there are at least $\sim 3000$ particles to resolve these halos.
For each simulation, we save all particles of 12 snapshots at $z=$\{3.00, 2.50, 2.00, 1.75, 1.50, 1.25, 1.00, 0.80, 0.50, 0.25, 0.10, 0.00\}.
Halos and subhalos are identified by \rockstar \cite{2013ApJ...762..109B} and \fof plus \subfind \cite{2001MNRAS.328..726S} on each snapshot.
In this work, we ignore the subhalos and only focus on the main halos.
Considering the measured HMF dependence on halo mass definition, we investigate three different halo catalogs: the \rockstar halos with an overdensity threshold of $\Delta=200$ times the mean matter (`cb') density ($M_{200m}$), the \rockstar halos with virial overdensity threshold ($M_{vir}$), and the \fof halos with overdensity criterion $\Delta=200\rho_{crit}$ ($M_{200c}$).
For each halo catalog, the cumulative HMF is calculated between $10^{10}$ and $10^{16}\, h^{-1}\Msun$ with a bin width of $0.1$ dex per decade, corresponding to $\mathrm{Nbin}=60$.

\subsection{Resolution Effect}
\label{sec:resolution}

The HMF measurements at the low-mass end are significantly affected by the limited mass resolution.
Thus, we need to carefully determine the minimum mass range to avoid the potential systematics.
In this section, we discuss the potential systematics of the cumulative HMF measurements due to the resolution effect.
We use three (CT0, CT3, and CT4) of the five convergence test simulations used in our previous work~\cite{Chen2025} to quantify the resolution effect.
They are generated at the Planck 2018 cosmology without massive neutrinos.
The volume of the simulation boxes is fixed at $250\,\mpch$ to reduce the computational resource consumption.
The first test simulation CT0 evolves $768^3$ particles, consistent with the mass resolution of \Kun simulations.
The particle loads of the other two test simulations are varied to $384^3$ (CT3) and $1536^3$ (CT4).

\begin{figure*}
    \centering
    \includegraphics[width=0.9\textwidth]{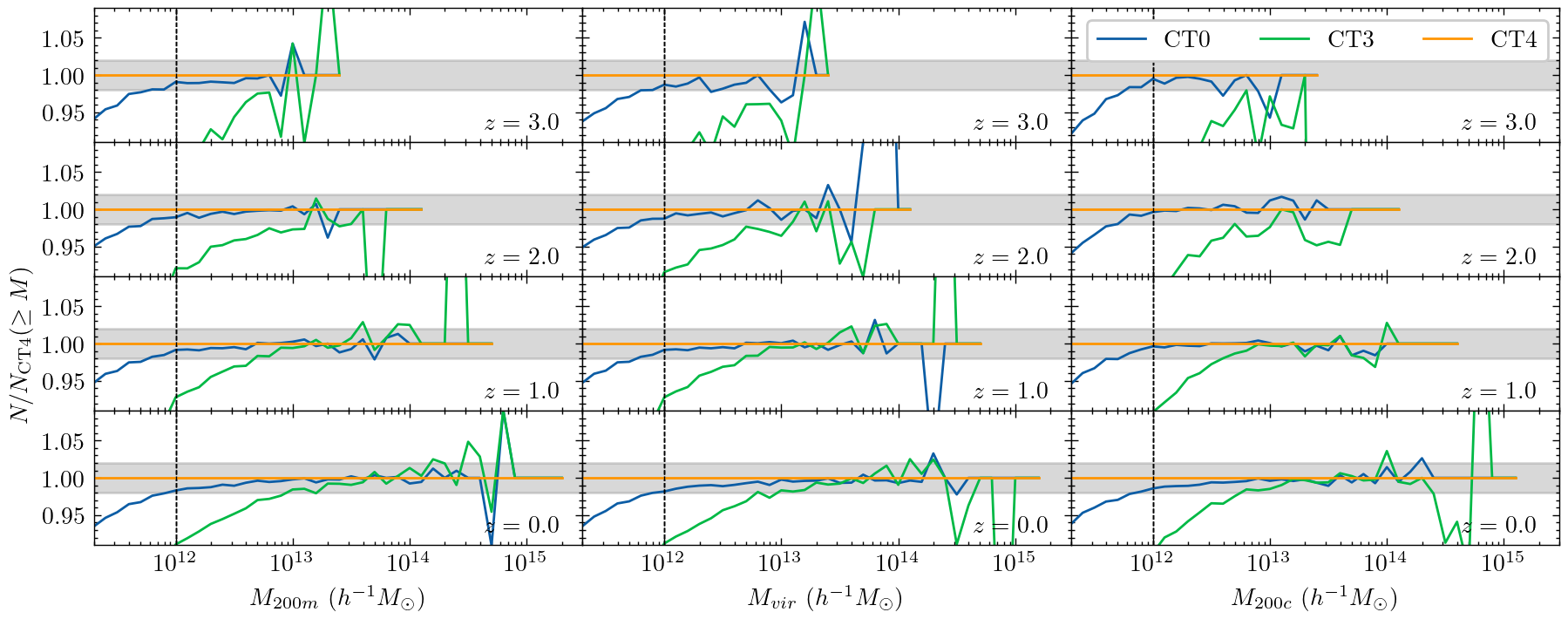}
    \caption{Comparison of the cumulative HMF between the CT0 simulation ($768^3$ particles, blue lines) and CT3 simulation ($384^3$ particles, green lines) to the CT4 simulation ($1536^3$ particles, orange lines).
    From left panel to right panel, we show the results of different halo definitions: $M_{200m}$, $M_{vir}$, and $M_{200c}$.
    The impact of mass resolution increases with increasing redshift (from bottom to top).
    The black dashed line indicates $10^{12}\,h^{-1}\Msun$ and gray region represents $\pm 2\%$ difference.
    }
    \label{fig:resolution}
\end{figure*}

The comparison of cumulative HMF between three test simulations and the one with the highest mass resolution is illustrated in Fig.~\ref{fig:resolution}.
The resolution effect increases with the increasing redshift and decreasing mass.
The halos with different mass definitions suffer similar impacts from the resolution effect.
The ratio of the cumulative HMFs between CT0 and CT4 is shown by the blue solid line.
From $z=3.0$ to $z=0.0$, the relative difference converges within $2\%$ for all halo mass definitions at $M\geq 10^{12}\,h^{-1}\Msun$, corresponding to $\gtrsim 300$ particles per halo.
This demonstrates that the impact of the resolution effect can be negligible for our fiducial simulation setup in the interested mass range.

Here, we do not attempt to use the empirical formula to correct the underestimation at the low-mass end.
Because these formulas are dependent on the specific simulation setup, e.g., the number of particles, the softening length, and even the halo finder algorithm.
Moreover, the mass range to perform the corrections is commonly assumed to be independent of cosmologies.
When the considered parameter space is broad, as in this work, the applicability of this assumption is unverified.
Fortunately, the measured HMF in our simulation setup has converged to $2\%$ accuracy at $M\geq 10^{12}\,h^{-1}\Msun$ for all redshifts and halo definitions.
Therefore, we focus on the emulation performance in this range.

\section{Emulator Construction}
\label{sec:emulation}

In this section, we demonstrate a generalized HMF emulator construction framework for different halo definitions.

\begin{figure*}
    \centering
    \includegraphics[width=0.95\textwidth]{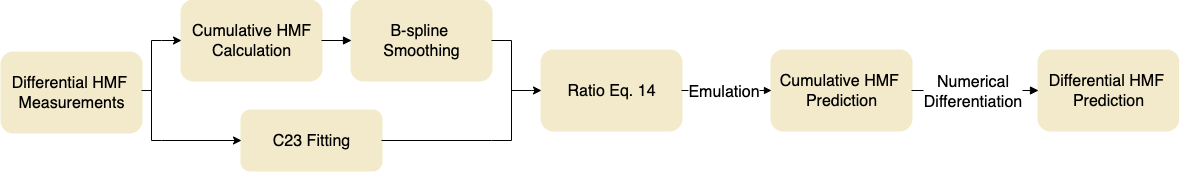}
    \caption{Flowchart for constructing the halo mass function emulator.}
    \label{fig:flowchart}
\end{figure*}

Directly emulating the measured HMF will introduce significant sample variance in final outputs.
To overcome this, a straightforward approach adopted in previous studies (e.g.,~\cite{2019ApJ...872...53M,2019ApJ...884...29N,2025JCAP...03..056S}) is to fit the measured HMF for each cosmological model through a specific parametrization such as Tinker08, and then to emulate the fitted parameters.
While we found that this parametrization is unable to fully capture the shape of the HMF and induce percent-level systematics within the extensive cosmological space of our work.
Alternatively, the Mira-Titan~\cite{2020ApJ...901....5B} and e-MANTIS~\cite{2024A&A...691A.323S} emulators utilized the more flexible B-splines to fit the HMF measurements under the $w_0w_a \mathrm{CDM}+\sum m_\nu$ cosmology and modified gravity model.
However, as a data-driven approach without a physical prior, the performance of B-spline fitting decreases fast when extrapolating to smaller or larger halo masses.
To take advantage of the physical prior introduced by the fitting formula but not be limited by the form of the fitting, we adopt a similar strategy as done in our previous work~\cite{Chen2025}.
We use a recently proposed fitting formula as the reference and emulate the difference between the measured HMF and the reference.
We briefly summarize three steps after measuring the cumulative HMF for all halo catalogs as follows:

\begin{itemize}
    \item Employ the B-spline decomposition to obtain the smoothed representation of the cumulative HMF. This procedure can effectively reduce the impact of sample variance in the high-mass end.
    \item In parallel, fit the multiplicity function $\nu f(\nu)$ in Eq.~\ref{eq:hmf} using a recently proposed fitting formula for all redshifts and cosmologies. The HMF calculated from the fitted $\nu f(\nu)$ can be used as a prior for the emulator and provides a more robust extrapolation at both high and low-mass ends.
    \item Emulate the ratio between the smoothed cumulative HMF and predictions from the fitted function using the combination of Gaussian process regression (GPR) and Principal Component Analysis (PCA).
\end{itemize}
Finally, the cumulative HMF emulator is constructed, and the corresponding differential HMF can be obtained by numerical differentiation.
The whole emulator construction flowchart, including the above steps, is presented in Fig.~\ref{fig:flowchart}.

\subsection{B-spline Decomposition}
\label{sec:Bspline}

Because of limited simulated volume, the HMF in simulations suffers a significant sample variance for high-mass halos.
Directly emulating the measurements will induce a significant bias and lead to worse emulation performance.
Inspired by the recent emulator works~\cite{2020ApJ...901....5B,2024A&A...691A.323S}, we employ the B-spline decomposition to obtain the smoothed cumulative HMF from simulations.
The measurements are decomposed into a set of B-splines as
\begin{equation}
    \log_{10} N(\geq M) = \sum_{i=0}^{n-1} c_{i}B_{i,k}(\log_{10} M) \ ,
    \label{eq:Bspline}
\end{equation}
where $c_{i}$ and $B_{i,k}$ are the coefficients and B-spline basis functions, respectively.
We calculate the B-spline basis elements via
\begin{equation}
    \begin{array}{l}
        B_{i, 0}(x)=1, \text { if } t_i \leq x<t_{i+1}, \text { otherwise 0}\ . \\
        B_{i, k}(x)=\frac{x-t_i}{t_{i+k}-t_i} B_{i, k-1}(x)+\frac{t_{i+k+1}-x}{t_{i+k+1}-t_{i+1}} B_{i+1, k-1}(x)\ .
    \end{array}
    \label{eq:Bspline-basis}
\end{equation}
Here, $t_{i}$ is an ensemble of knots with a total number of $n+k+1$,
where $n$ is the number of B-splines and $k$ represents the degree of B-splines.
We fix the spline degree $k=2$, and the knots are chosen in the logarithmic mass space with equal separation of $\Delta \log_{10}M =0.25$ in this work (e.g.,~\cite{2020ApJ...901....5B,2024A&A...691A.323S}).
To obtain a better estimation at the low-mass end of emulation ($M\sim 10^{12}\,h^{-1}\Msun$), the minimum mass in the B-spline decomposition is chosen at $10^{11} \,h^{-1}\Msun$, an order of magnitude lower than $10^{12}\,h^{-1}\Msun$.
We have checked that this extended range does not affect the performance of the B-spline decomposition procedure.

Due to the broad cosmological space considered in this work, the maximum halo mass in each snapshot varies rapidly across different models.
While the latter GPR needs a uniform high-mass end.
A conservative (relatively high mass) cut may introduce extra bias to the emulator, while a tight (relatively low mass) cut will discard the information about the high-mass end.
It is essential to determine a reasonable maximum halo mass for all cosmologies to balance the performance and the applicable range of the emulator.
We set the maximum halo mass using the same procedure in~\cite{2024A&A...691A.323S},
\begin{equation}
    M_{\max } (z)=\alpha \times 10^{\left\langle\log_{10} M_{\max , i} (z)\right\rangle_{\theta_i}}\, ,
    \label{eq:max-mass}
\end{equation}
where $M_{\max , i}$ represents the maximum halo mass under a given cosmology $\theta_i$.
Here, the average is adopted across different cosmologies.
We have checked that $\alpha = 0.5$ for all halo definitions and redshifts is a good choice.
At the same time, we only utilize the data points with the minimum number of halos $20$ to avoid biased B-spline results at the high-mass end, and the Poisson noise is also considered as the weights during the fitting.

\begin{figure}[H]
    \centering
    \includegraphics[width=0.4\textwidth]{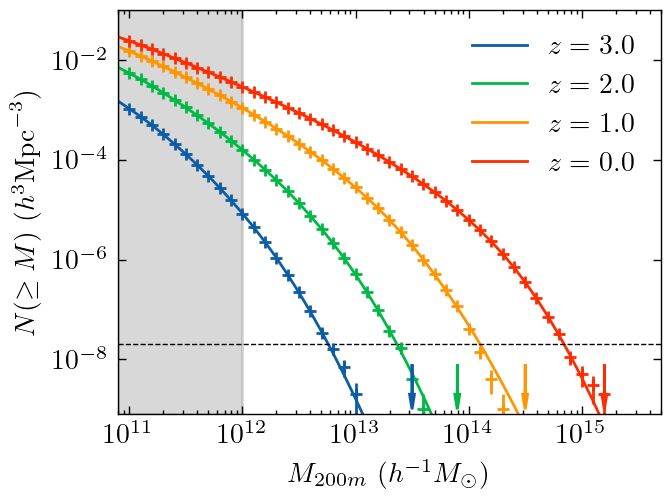}
    \caption{The measured cumulative HMF and the B-spline decomposition results for the c0001 cosmology.
    The vertical error bars represent the Poisson noise for each data point.
    The horizontal dashed line indicates the minimum number of halos $20$.
    Different colors represent results at different redshifts.
    And the arrow with the same color exhibits the mass threshold in Eq.~\ref{eq:max-mass} at the corresponding redshift.
    The gray region indicates the range suspicious to systematics, $M_{200m}<10^{12}\,h^{-1}\Msun$.}
    \label{fig:b-spline}
\end{figure}

In Fig.~\ref{fig:b-spline}, we choose the c0001 cosmology as an example to illustrate the robustness of the B-spline decomposition,
because c0001 has $\sigma_8=0.6059$ and thus less massive halos and larger sample variance.
Only the data points above the black dashed line are utilized for the B-spline decomposition.
The colored arrows represent the mass thresholds at four redshifts driven from Eq.~\ref{eq:max-mass}.
Considering the percent-level resolution effect at $M<10^{12}\,h^{-1}\Msun$ (gray shadow) as discussed in Sec.~\ref{sec:resolution}, we do not recommend utilizing the results in this area.
Though the B-spline decomposition can describe the cumulative HMF well at the low-mass end.

\begin{figure*}
    \centering
    \includegraphics[width=0.9\textwidth]{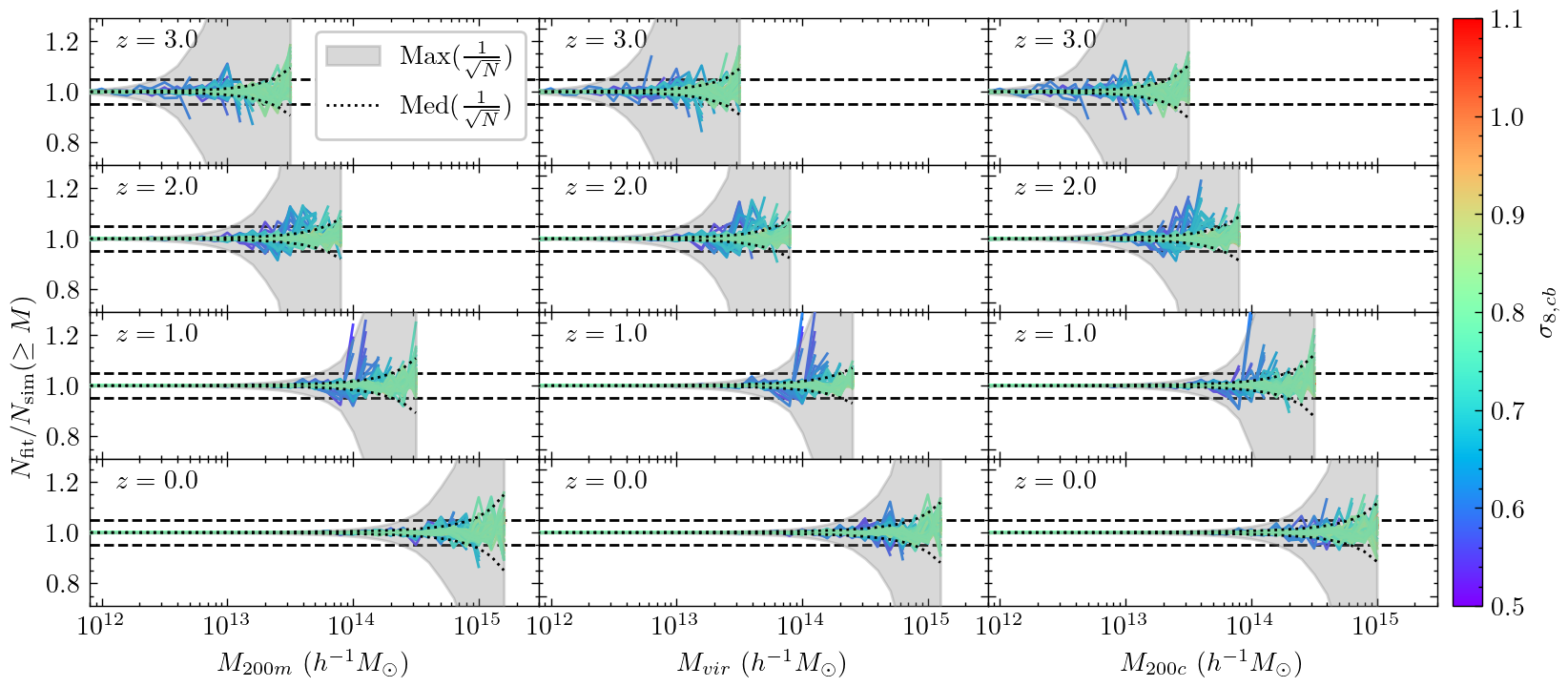}
    \caption{The ratios between the B-spline decomposition results and measurements from simulations for all cosmologies and three mass definitions: $M_{200m}$ (left panels), $M_{vir}$ (middle panels), and $M_{200c}$ (right panels).
    The results of $z=3.0,\,2.0,\,1.0,\,0.0$ are shown from top to bottom.
    The black dotted line and gray region represent the median and maximum Poisson noise over all cosmologies, respectively.
    The black dashed line indicates $\pm 5\%$ difference.
    Data with the number of halos smaller than $20$ (not used in the B-spline decomposition fitting) are removed.
    }
    \label{fig:b-spline-error}
\end{figure*}

In practice, we adopt \texttt{LSQUnivariateSpline} in the Python package \textsc{scipy}~\cite{2020SciPy-NMeth} to calculate the B-spline coefficients for each halo catalog.
The coefficients are obtained through minimizing the weighted least-squares error between the measured and smoothed cumulative HMF.
The errors induced by this smooth procedure for all halo definitions are illustrated in Fig.~\ref{fig:b-spline-error}.
We remove the points whose number of halos is smaller than $20$ for better visualization.
The results for different cosmologies are sorted according to their $\sigma_{8,cb}$ values.
A similar performance is observed for different halo definitions.
At low-mass ends, this method exhibits excellent agreement with the cumulative HMF measured from simulations.
For the high-mass halo, the relative difference for most cosmologies is within the median Poisson noise shown with the black dotted lines.
The results with low clustering amplitudes ($\sigma_{8,cb} \lesssim 0.7$) exhibit slightly poor performance.
While the errors are still within the maximum Poisson noise (gray shadow) across different cosmologies.
Note that the difference for cosmologies with high $\sigma_{8,cb}$ is also within the median Poisson noise, but the red lines are all hidden behind the green lines.
This demonstrates that the observed errors in this procedure are mainly contributed by the Poisson noise.

\subsection{Multiplicity Fitting Formula}
\label{sec:multiplicity-fit}

In our first work~\cite{Chen2025}, an accurate theoretical prediction can help improve the final emulator accuracy significantly.
Thus, we employ a flexible fitting formula to describe the smoothed measurements in this section.
The fitted results are taken as the denominator to scale the dynamical range of the quantity to emulate.
After testing various HMF parameterizations, we find the formula in the Euclid preparation~XXIV~\cite{2023A&A...671A.100E} (hereafter Castro23, also C23 in equations) can describe our measurements well.
The differential HMF can be expressed as:
\begin{equation}
    \nu f(\nu)=A(p, q) \sqrt{\frac{2 a \nu^2}{\pi}} \mathrm{e}^{-a \nu^2 / 2}\left(1+\frac{1}{\left(a \nu^2\right)^p}\right)(\nu \sqrt{a})^{q-1}\, ,
    \label{eq:C23-hmf}
\end{equation}
where $A(p,q)$ is the normalization factor with the form:
\begin{equation}
    A(p, q)=\left\{\frac{2^{-1 / 2-p+q / 2}}{\sqrt{\pi}}\left[2^p \Gamma\left(\frac{q}{2}\right)+\Gamma\left(-p+\frac{q}{2}\right)\right]\right\}^{-1}\, .
    \label{eq:C23-norm}
\end{equation}
The fitting parameters $a, p, q$ capture the cosmological and redshift dependence of HMF.
We calculate them from linear theory through
\begin{equation}
\begin{array}{cc}
a =& a_R \Omega_{\mathrm{m}}^{a_z}(z)\, ,\\
p =& p_1+p_2\left(\frac{\mathrm{~d} \ln \sigma}{\mathrm{~d} \ln R}+0.5\right)\, ,\\
q =& q_R \Omega_{\mathrm{m}}^{q_z}(z)\, , \\
a_R =& a_1+a_2\left(\frac{\mathrm{~d} \ln \sigma}{\mathrm{~d} \ln R}+0.6125\right)^2\, , \\
q_R=&q_1+q_2\left(\frac{\mathrm{~d} \ln \sigma}{\mathrm{~d} \ln R}+0.5\right)\, .
\end{array}
\label{eq:paramert-C23hmf}
\end{equation}

The $\sigma$ and $R$ are defined in Eq.~\ref{eq:sigma}.
We limit $M \geq 10^{12}\,h^{-1}\Msun$ for all halo definitions during the fitting process.
For a given set of parameters $\boldsymbol{\theta}$, we calculate the halo number counts $N_{\alpha i}(\boldsymbol{\theta}, z_{\alpha})$ by integrating Eq.~\ref{eq:hmf} at redshift $z_{\alpha}$ in $i$-th mass bin $[M_{i},\,M_{i+1}]$ and minimize the following likelihood
\begin{equation}
\label{eq:c23-likelihood}
\ln \mathcal{L} = \begin{cases}\sum_{\alpha i}(N_{\alpha i}^{\operatorname{sim}} \ln \left(\frac{N_{\alpha i}(\boldsymbol{\theta}, z_{\alpha})}{N_{\alpha i}^{\text {sim }}}\right)-N_{\alpha i}(\boldsymbol{\theta}, z_{\alpha})+N_{\alpha i}^{\operatorname{sim}}) ,
\\ \hfill N_{\alpha i}^{\mathrm{sim}} \leq 25, \\ 
\frac{1}{2} \ln \left(2 \pi \sigma^2\right)+\sum_{\alpha i}\frac{\left(N_{\alpha i}(\boldsymbol{\theta}, z_{\alpha})-N_{\alpha i}^{\text {sim }}\right)^2}{2 \sigma^2}, \\ \hfill N_{\alpha i}^{\mathrm{sim}}>25.\end{cases} 
\end{equation}
Here, $N^{\mathrm{sim}}_{\alpha i}$ denotes the number count of halos at redshift $z_{\alpha}$ and $i$-th mass bin in simulations. 
And $\sigma=\sqrt{N_{\alpha i}^{\operatorname{sim}}+\sigma_{\mathrm{sys}}^2}$ represents the standard error caused by the combination of Poisson noise and an extra noise term, such as round-off errors.
We assume $\sigma_{\rm sys}=0.005N^{\mathrm{sim}}_{\alpha i}$ as in~\cite{2023A&A...671A.100E}.
The final likelihood is given by the sum of two expressions in Eq.~\ref{eq:c23-likelihood} across all 129 cosmologies.
The fitted precision decreases as $z$ decreases, reaching $\sim 5\%$ for $M\in [10^{12}, 10^{14}]\,h^{-1}\Msun$ at $z=0.0$.
The calibrated parameters are concluded in Tab.~\ref{tab:fitting}.
The performance of a more complex model in a recent updated work~\cite{C25} is also tested, but no significant improvement is observed.
Thus, we integrate the calibrated Castro23 model as the denominator in Sec.~\ref{sec:emuhmf}.

\begin{table*}
    \caption{Calibrated parameters in Eq.~\ref{eq:C23-hmf} for three kinds of halo catalogs in \Kun simulation suite.}
    \centering
    \begin{tabular}{ccccccccc}
    \toprule
    Mass Def. & $a_1$ & $a_2$ & $a_z$ & $p_1$ & $p_2$ & $q_1$ & $q_2$ & $q_z$ \\
    \midrule
    \rockstar $M_{200m}$ & 0.7728 & 0.3686 & -0.0321 & -0.4608 & -0.5342 & 0.3734 & -0.3232 & -0.1920 \\
    \rockstar $M_{vir}$ & 0.7645 & 0.4171 & -0.0618 & -0.4686 & -0.6511 & 0.3883 & -0.3696 & -0.0069 \\
    \fof $M_{200c}$ & 0.7660 & 0.3854 & -0.1423 & -0.5236 & -0.7343 & 0.3208 & -0.4046 & 0.1527 \\
    \bottomrule
    \end{tabular}
    \label{tab:fitting}
\end{table*}

\subsection{Emulation}
\label{sec:emuhmf}

The last step of emulator construction is to emulate the given quantity using the combination of PCA and GPR.
Similar to the previous work~\cite{Chen2025}, we construct the emulator based on the ratio between the smoothed cumulative HMF and the integration of calibrated Castro23 HMF
\begin{equation}
    R_j(\geq M_j,z) = \frac{N_j(\geq M_j,z)}{N_{j, C23}(\geq M_j, z)}\, ,
    \label{eq:R_m}
\end{equation}
where $j$ represents three different halo definitions: $\{ M_{200m}$, $M_{vir}$, and $M_{200c} \}$.
After applying the different mass cuts at different redshifts, we obtain $N_{d} = 425$ data points for each cosmology.
Then PCA technique is utilized to reduce the cross-correlation and dimensionality of the training data.
We find that taking the first 10 principal components ($N_\mathrm{PCA}=10$) is sufficient to recover the smoothed data with a difference within the median Poisson noise.
At last, we employ GPR to perform the interpolation in the eight-dimensional cosmological parameter space.
Different from our first work, a combination of a constant kernel and the flexible Marten-$5/2$ kernel is adopted in this work.
Other detailed descriptions can be found in Sec.~4.2 of the previous work~\cite{Chen2025}.
The PCA and GPR in the Python package \textsc{scikit-learn} \cite{scikit-learn} are utilized as the practical implementation.
We also validate that there are negligible accuracy improvements from the emulation without PCA, i.e., emulating $N_{d} = 425$ data points at different halo masses and redshifts separately.
While the time consumption of the training and prediction process is $\sim40$ times that of the version with PCA.
Thus, we utilize the PCA mainly to obtain the fast emulation of HMF ($\sim0.3\,\mathrm{s}$ per cosmology).

\subsection{Leave-One-Out Validation}
\label{sec:loo}

\begin{figure*}
    \centering
    \includegraphics[width=0.9\textwidth]{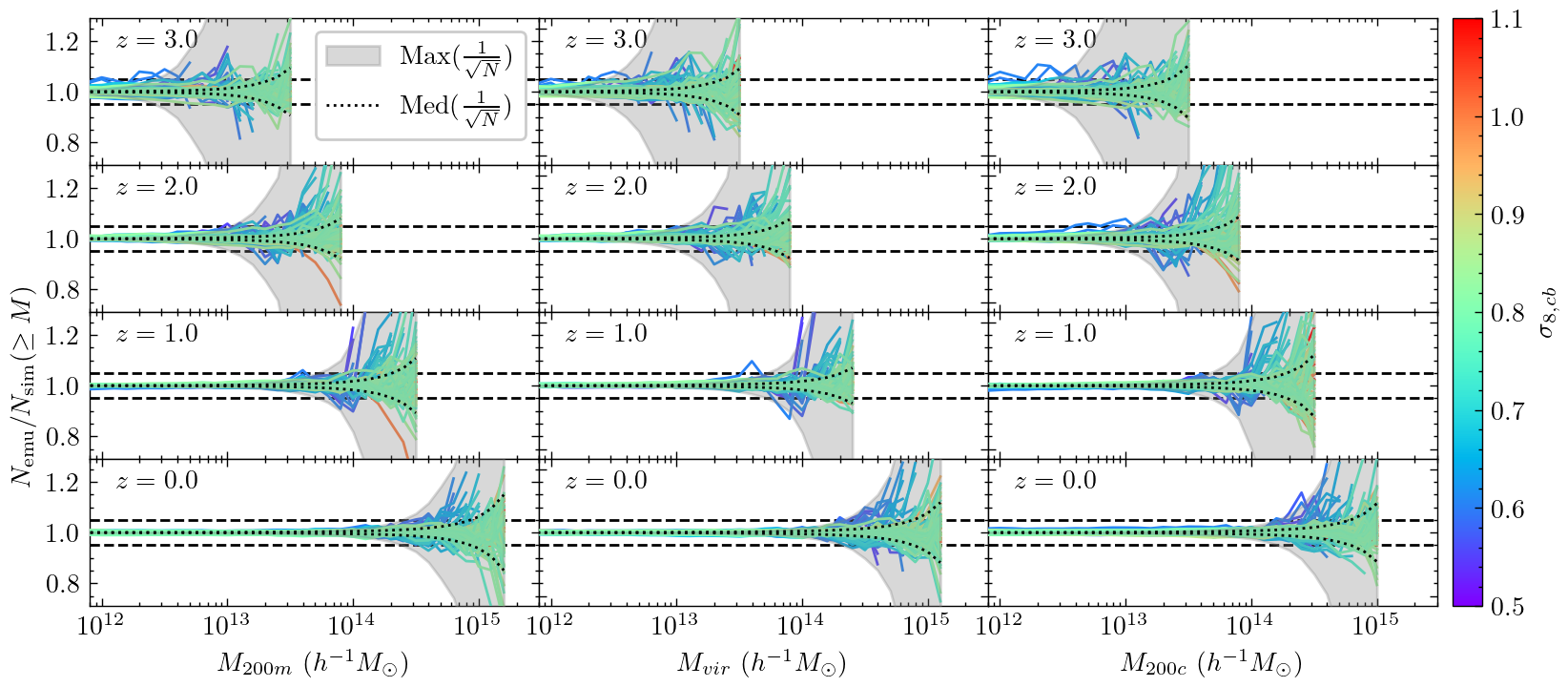}
    \caption{Similar to Fig.~\ref{fig:b-spline-error}, but for the comparison between the LOO predictions and the simulated cumulative HMF.}
    \label{fig:emu-error}
\end{figure*}

\begin{figure*}
    \centering
    \includegraphics[width=0.9\textwidth]{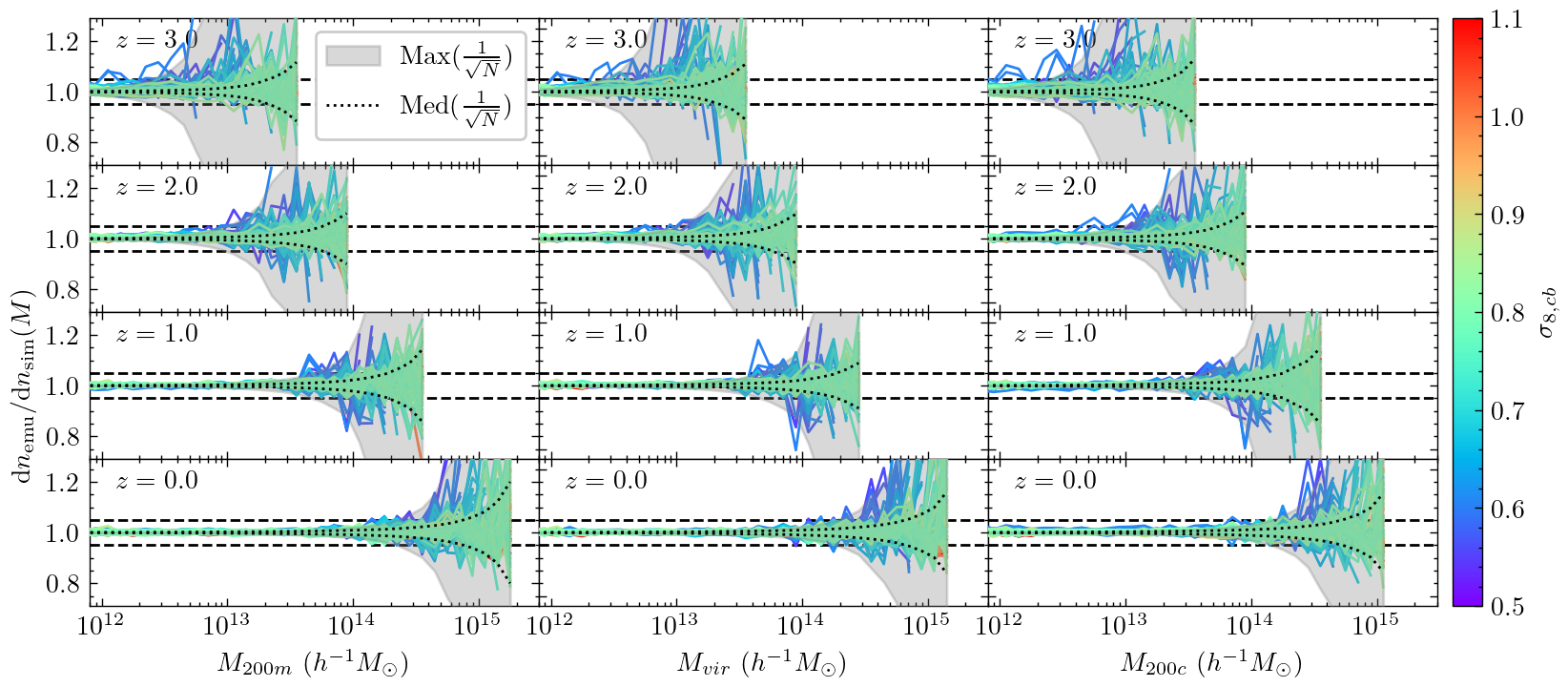}
    \caption{Similar to Fig.~\ref{fig:emu-error}, but comparing the LOO predictions and the simulated differential HMF.}
    \label{fig:emu-dndlgM-error}
\end{figure*}

The remaining task is to quantify the emulator performance.
We employ the leave-one-out (LOO) cross-validation method to validate our emulator accuracy.
Each time, we choose only one simulation as the validation set, while the rest of the 128 simulations are taken as the training set.
Thus, we can obtain a robust estimate of the emulator performance across the whole parameter space.
The ratios between the emulator prediction and simulation results are illustrated in Fig.~\ref{fig:emu-error}.
For halos with mass $M\lesssim 10^{13}\, (10^{14})\,h^{-1}\Msun$ at $z\leq1.0$, the relative difference is within $2\% \, (5\%)$.
The slightly poor performance of the cluster halo mass is caused by the large statistical errors in the training data.
In the hierarchical structure formation scenario, as redshift increases, the mass of the most massive halo decreases quickly, leading to more Poisson noise.
The relative difference for the samples with slightly higher clustering amplitude $\sigma_{8,cb}\gtrsim 0.7$ (green and red lines) is within the median Poisson noise of all cosmologies (black dotted lines).
For the cosmology with a lower $\sigma_{8,cb}$ value, there are fewer halos in the limited simulation volume, which is equivalent to the results at high redshifts to some degree.
Thus, more statistical errors for the blue lines are observed.
In conclusion, the observed difference between LOO predictions and simulated results is mainly caused by statistical fluctuations.

Another interesting thing is that the accuracy of our emulators for three different halo mass definitions is at the same level from left to right panels in Fig.~\ref{fig:emu-error}.
This demonstrates that the generalized emulation framework in this work is robust for different halo definitions and halo finding algorithms.
For some more physical halo boundaries, e.g., the splashback radius (e.g.,~\cite{1984ApJ...281....1F,1985ApJS...58...39B,2014JCAP...11..019A,2014ApJ...789....1D}) and depletion radius (e.g.,~\cite{2021MNRAS.503.4250F,2022MNRAS.513.4754F,2023ApJ...953...37G,2023MNRAS.525.2489Z,2025ApJ...979...55Z}), there are no corresponding emulators to provide the mass function prediction.
The generalization of our \texttt{CSST Emulator} makes this extension easier once the corresponding halo catalogs are identified.

To compare with the existing emulators and fitting formulae, we also illustrate the relative difference between the LOO predictions of emulators and the measured differential HMF in Fig.~\ref{fig:emu-dndlgM-error}.
The differential HMF predictions are derived from numerical differentiation, and the binning effect has been accounted for through a similar integration procedure in Sec.~\ref{sec:binning}.
The overall relative difference is slightly larger than the results of the cumulative form, which is led by the larger statistical errors of the differential HMF.
While the observed errors for most cosmologies are still within the Poisson noise.
The trends with variations of redshifts, halo mass, and halo definitions are all consistent with the cumulative HMF, confirming the previous findings.

\section{Validation and Application}
\label{sec:validation}

\subsection{Comparison with Other Tools}

\begin{figure}[H]
    \centering
    \includegraphics[width=0.4\textwidth]{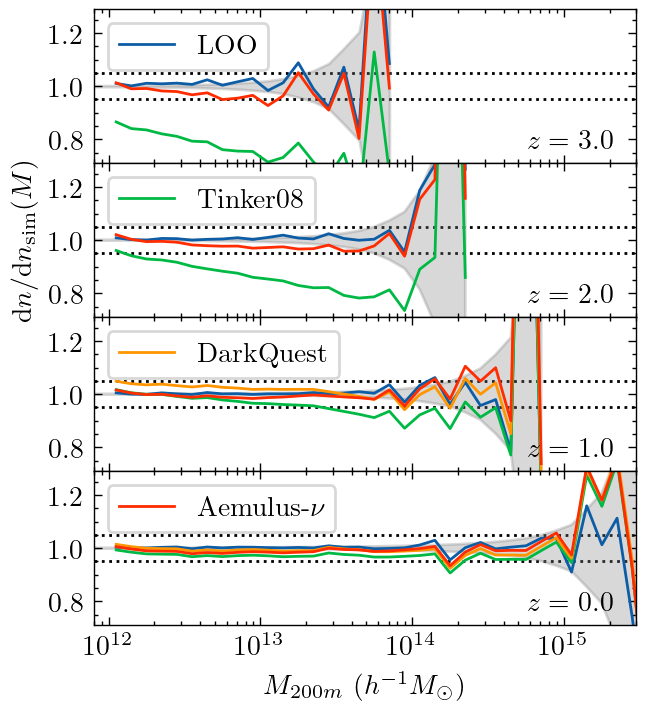}
    \caption{Comparison of different halo mass function predictions with the simulation for the fiducial model c0000 from redshift $z=3.0$ to $z=0.0$.
    The LOO (blue line) represents the \texttt{CSST Emulator} prediction using the LOO method detailed in Sec.~\ref{sec:loo}.
    The yellow solid lines show the DarkQuest results only at $z=0.0$ and $1.0$ due to the redshift limit of this emulator, $[0.0, 1.48]$.
    The predictions of the fitting formula from Tinker08 and the recent Aemulus-$\nu$ emulator are illustrated by green and red lines, respectively.
    The gray region indicates the statistical error of the simulation.
    The black dotted lines represent $5\%$ difference.
    }
    \label{fig:c0000-hmf}
\end{figure}

To further illustrate the validity of our emulator, we compare differential HMF predictions from different tools with the measurements from our fiducial simulation c0000 in Fig.~\ref{fig:c0000-hmf}.
The accuracy of \texttt{CSST Emulator} is represented by the LOO result, which was predicted by the emulator training only using simulations c0001-c0128 for a fair comparison.
The LOO predictions exhibit excellent agreement with the simulation measurements, as presented by blue lines.
The Tinker08 prediction (green) is calculated from the fitting formula by ignoring the neutrino clustering in the matter power spectrum as previous work~\cite{2014JCAP...02..049C}.
The difference increases as the redshifts, due to the increased non-universality of the HMF under the Tinker08 parameterization at high redshifts.
This finding is consistent with other works (e.g.,~\cite{2025JCAP...03..056S}).
The predictions from the Aemulus-$\nu$\footnote{\url{https://github.com/DelonShen/aemulusnu_hmf}} and DarkQuest\footnote{\url{https://dark-emulator.readthedocs.io/}} emulators converge with the simulation for $z \leq 1.0$, within the statistical error (gray region).
The slightly large deviation for the Aemulus-$\nu$ emulator at high redshifts is probably due to the difference in the configuration of the training simulations.
Note that the Aemulus-$\nu$ article only validates the emulator accuracy for $M_{200m} \geq 10^{13}\, h^{-1}\Msun$ although the published tool can provide predictions below this mass threshold.

In summary, the fitting formula and the emulators can obtain converged results at $z=0.0$.
While the accuracy of the fitting formula decreases fast at higher redshifts due to the increased non-universality (mainly the redshift dependence).
The predictions from all emulators agree well with the simulation measurements.

\subsection{Forecast for CSST Cluster Cosmology }
\label{sec:mcmc}

In this section, we forecast the parameter constraining power from cluster number counts for the upcoming CSST survey to exhibit the necessity of the HMF emulator.
Following previous CSST cluster cosmology forecast works~\cite{2023MNRAS.519.1132M,2023RAA....23d5011Z}, we build a mock halo catalog by assuming the mass distribution follows the smoothed HMF under our fiducial Planck-2018 cosmology, c0000.
The halo mass is defined as the virial mass $M_{vir}$, as this mass definition is adopted in the fitting formula from Castro23.
The expected number of clusters in the $i$-th mass bin and $\alpha$-th redshift bin is calculated by
\begin{equation}
N_{\alpha i}=\int_{\Delta z_\alpha} \mathrm{d} z \frac{\mathrm{~d} V}{\mathrm{~d} z} \int_{\Delta M_i} \mathrm{~d}\ln M \frac{\mathrm{~d} n}{\mathrm{~d}\ln M}(M, z)\, ,
\label{eq:cluster-number}
\end{equation}
where,
\begin{equation}
\frac{d V}{d z}=\Delta \Omega\left(\frac{\pi}{180}\right)^2 \frac{c}{H(z)}\left(\int_0^z d z^{\prime} \frac{c}{H\left(z^{\prime}\right)}\right)^2\, .
\label{eq:dvdz}
\end{equation}
Here, $\Delta \Omega = 17,500$ represents the survey area in square degrees.
We consider the cluster halo mass range $M_{vir}\in [10^{14},\,10^{16}]\,h^{-1}\Msun$ for $0\leq z \leq 1.5$.
The redshift range is equally divided by 10 bins with $\Delta z=0.15$.
We utilize the Monte Carlo Markov Chain method using emcee\footnote{\url{https://emcee.readthedocs.io/en/stable/}}, through maximizing
\begin{equation}
\label{eq:mcmc}
\ln \mathcal{L} = \sum_{\alpha i} (N_{\alpha i}-N^{\rm model}_{\alpha i}) \mathrm{Cov^{-1}} (N_{\alpha i}-N^{\rm model}_{\alpha i})^{T}\, .
\end{equation}
For the covariance, only the shot noise and a constant systematic error that incorporates potential systematics in the mass estimation are considered in this work (also see e.g.,~\cite{2019MNRAS.488.4779C,2023MNRAS.519.1132M}).
\begin{equation}
\label{eq:cov}
\mathrm{Cov} = \mathrm{Cov^{SN}} + \mathrm{Cov^{sys}}\,,
\end{equation}
where
\begin{equation}
\label{eq:covSN}
\mathrm{Cov^{SN}}_{\alpha \beta i j} = N_{\alpha i}\delta_{\alpha \beta}\delta_{i j}\, .
\end{equation}
And we assume $\mathrm{Cov^{sys}}$ is $30\%$ of the shot noise term.
We vary five cosmological parameters, $\{ \Omega_m,\,\sigma_8,\,\Omega_b,\,n_s,\,H_{0} \}$ during this sampling.
While only the 2D posterior distributions of $\Omega_m$ and $\sigma_8$ are illustrated due to the weak constraint power on other parameters.

\begin{figure}[H]
    \centering
    \includegraphics[width=0.4\textwidth]{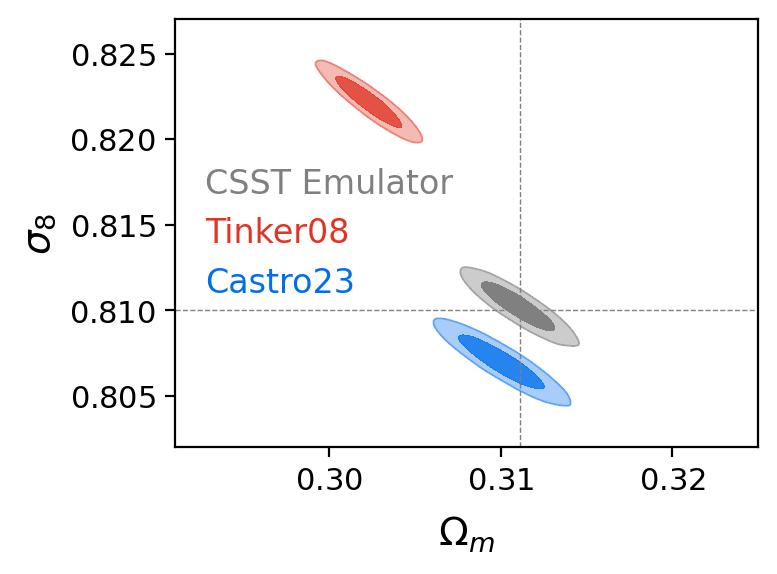}
    \caption{Cosmological constraints on $\Omega_m$ and $\sigma_8$ from the CSST-like survey for three different theoretical predictions, this work (gray), Tinker08 (red), and Castro23 (blue).
    }
    \label{fig:mcmc}
\end{figure}

We compare the performance of three different theoretical models, \texttt{CSST Emulator}, Tinker08, and Cartro23 in Fig.~\ref{fig:mcmc}.
Note that we have integrated all three models into our \texttt{CSST Emulator} package for a fair and convenient comparison.
Our emulator can obtain tight constraints with $\sigma_{\Omega_m}=0.0014$ and $\sigma_{\sigma_8}=0.0010$ without systematic bias under the $\Lambda\mathrm{CDM}$ framework.
However, the predictions from both fitting formulas exhibit significant systematic bias.
Especially for the Tinker08 model, both cosmological parameters deviate far from fiducial values.
On one hand, the significant deviations are caused by the poor performance at high redshifts indicated by Fig.~\ref{fig:c0000-hmf}.
On the other hand, the Tinker08 model was not specifically designed for $M_{vir}$.
The parameters for virial mass in the Tinker08 formula are obtained from interpolation over the overdensity threshold parameter $\Delta$.
On the contrary, our generalized emulator framework makes it easy to build an accurate prediction tool for any halo mass definition.
The recently proposed Castro23 fitting formula was calibrated using the \rockstar halo catalog (defined by $M_{vir}$) in their work~\cite{2023A&A...671A.100E}.
It can obtain more accurate constraints compared to Tinker08.
But $\sim 3\sigma$ deviation on the clustering amplitude $\sigma_8$ is also observed.
This application test quantifies the parameter constraint power and demonstrates the advantage of the emulator approach.
Note that we only include the systematics induced from mass estimation through an increased covariance in this work, which is very optimistic compared to the real observation.
We leave the more realistic and detailed analysis to future work.

In summary, the upcoming photometric and spectroscopic survey of CSST will constrain the matter density $\Omega_m$ and $\sigma_8$ with unprecedented accuracy.
However, the unbiased estimation is not ensured by the traditional fitting formula tools.
They are frequently built based on the universality of HMF, which has been proven to break down for high redshifts, different halo mass definitions, and high-dimensional cosmological space.
Emulation is a potential approach to overcome this.

\section{Conclusion and Discussions}
\label{sec:conc}

The forthcoming CSST observation will obtain unprecedented photometric and slitless spectroscopic data, which provides a new opportunity to study the expansion history and structure growth of the Universe.
A more accurate theoretical prediction is essential to match the reduced statistical errors of the new observational data.
For the theoretical preparation of CSST cosmology surveys, we generated the \Kun simulation suite, consisting 129 high-resolution simulations with particle mass resolution $2.87 \frac{\Omega_\mathrm{cb}}{0.3} \times 10^{9}\, h^{-1}\Msun$ across a broad range of cosmologies under the $w_0w_a \mathrm{CDM}+\sum m_\nu$ model.
Based on it, the $1\%$ percent accurate nonlinear matter power spectrum emulator for $k\leq 10\,\mpch$ has been constructed in our first work~\cite{Chen2025}.
In this work, the second article in the \texttt{CSST Emulator} series, we extend our \texttt{CSST Emulator} to provide accurate cumulative HMF and differential HMF predictions for various cosmological probes, e.g., cluster counts, galaxy clustering, and weak lensing.

In this work, we first investigate the impact of different binning schemes on the differential HMF.
We find the errors from binning effect reach $5\%$ at $M_{200m} \sim 3\times10^{15} \, h^{-1}\Msun$ and $z=0.0$ with bin width $0.1$ dex.
The binning effect measured from the averaged HMF of 100 Quijote-HR simulations can be well described by the predicted bias using the Tinker08 model.
This indicates that our understanding of the binning effect is correct.
Moreover, this inspires us to utilize the cumulative form of HMF to avoid the influence of different mass binning schemes.

Then, we demonstrate a generalized framework to emulate the cumulative HMF for three different halo definitions: $M_{200m}$, $M_{vir}$, and $M_{200c}$.
The emulation procedure is based on the ratio of the smoothed cumulative HMF and the integrated theoretical prediction.
The B-spline decomposition provides a robust and effective way to smooth the measurements from simulations, as shown in Fig.~\ref{fig:b-spline-error}..
To gain as much as possible cosmological dependence from theoretical predictions, we re-calibrate eight free parameters in the Castro23 model (Eqs.~\ref{eq:C23-hmf}-\ref{eq:paramert-C23hmf}).
The calibrated parameters are concluded in Table~\ref{tab:fitting}.
The ratios of the B-spline results and the integration of calibrated Castro23 HMF are emulated using the combination of PCA and GPR.
We quantify the emulator performance using the LOO cross-validation method, which is illustrated in Fig.~\ref{fig:emu-error}.
For the low-mass halo ($M\leq 10^{13}\,h^{-1}\Msun$) at $z\leq 2.0$, errors from our predictions are roughly $1\sim 2\%$.
The accuracy at cluster mass $M\gtrsim 10^{14}\, (10^{15}) \,h^{-1}\Msun$ are within $5\%$ ($10\%$).
For all cases, the relative difference between the emulator prediction and simulations is within the median Poisson noise for cosmologies with higher $\sigma_{8,cb}$ ($\gtrsim 0.7$).
The poor results from cosmologies with small $\sigma_{8,cb}$ are still comparable with the statistical errors of the training data.
The prediction accuracy for different halo definitions is similar, indicating that our generalized framework is robust to emulate the cumulative HMF for various halo definitions and halo finders.
The accuracy of the differential HMF is illustrated in Fig.~\ref{fig:emu-dndlgM-error}, showing consistent results with the cumulative one.
To further validate our emulator accuracy, we compare the LOO prediction with other theoretical tools in Fig.~\ref{fig:c0000-hmf} and find consistent results for all emulators.
The Tinker08 fitting formula exhibits significant deviation from our c0000 simulation measurements at high redshifts.
Moreover, we forecast the cosmological constraints for the upcoming CSST observation for three different theoretical models.
Both traditional fitting formulas introduce significant systematics on $\Omega_m$ and $\sigma_8$ compared to \texttt{CSST Emulator} results.
This suggests that it is necessary to develop an accurate HMF emulator for unbiased constraints from cluster abundance in the Stage-IV survey, such as CSST.

The HMF emulator has been integrated into the published \texttt{CSST Emulator} at \url{https://github.com/czymh/csstemu}.
The user-friendly package is designed exquisitely and only depends on \textsc{numpy} \cite{2020Natur.585..357H} and \textsc{scipy} \cite{2020SciPy-NMeth}.
This tool is documented at \url{https://csst-emulator.readthedocs.io/}, and more useful statistics are under active development.

For the next step, we plan to construct the halo bias emulators based on \Kun simulations, which are useful for the galaxy clustering analysis.
With the accurate predictions of halo mass function and halo bias, we can exploit the full cosmological information from cluster abundance and clustering to provide tight constraints on $\Omega_m$ and $\sigma_8$.
Moreover, our halo mass function can provide accurate number counts for halo masses down to $10^{12}\,h^{-1}\Msun$.
We can construct the halo clustering emulators and utilize the theoretical galaxy-halo connection to obtain the galaxy correlation and galaxy-galaxy lensing cross-correlation functions (e.g.,~\cite{2019ApJ...884...29N}).
This will provide a more comprehensive picture of the large-scale structure and the evolution of the Universe.

\Acknowledgements{
This work was supported by the National Key R\&D Program of China (No. 2023YFA1607800, 2023YFA1607801, 2023YFA1607802), the National Science Foundation of China (Grant Nos. 12273020, 12133006), the China Manned Space Project with No. CMS-CSST-2021-A03 and CMS-CSST-2025-A04, the ``111'' Project of the Ministry of Education under grant No. B20019,
and the sponsorship from Yangyang Development Fund.
The \textsc{Kun} simulation suite is run on 
Kunshan Computing Center.
The analysis is performed on the Gravity
Supercomputer at the Department of Astronomy, Shanghai Jiao Tong University, and the $\pi \,2.0$ cluster supported by the Center for High Performance Computing at Shanghai Jiao Tong University.
}

\InterestConflict{The authors declare that they have no conflict of interest.}


\bibliographystyle{scpma}
\bibliography{biblio.bib}

\begin{thebibliography}{100}
\providecommand{\url}[1]{\texttt{#1}}
\providecommand{\urlprefix}{URL }
\providecommand{\doi}[1]{doi:~\href{http://doi.org/#1}{\nolinkurl{#1}}}
\providecommand{\arXiv}[1]{\href{https://arxiv.org/abs/#1}{\nolinkurl{https://arxiv.org/abs/#1}}}
\providecommand{\eprint}[1]{\href{http://arxiv.org/abs/#1}{\nolinkurl{#1}}}

\bibitem{2001ApJ...561...13B}
S.~{Borgani}, P.~{Rosati}, P.~{Tozzi}, S.~A. {Stanford}, P.~R. {Eisenhardt}, C.~{Lidman} et~al., \apj \textbf{561}, 13 (2001), arXiv: \eprint{astro-ph/0106428}.

\bibitem{2003A&A...398..867S}
P.~{Schuecker}, H.~{B{\"o}hringer}, C.~A. {Collins}, and L.~{Guzzo}, \aap \textbf{398}, 867 (2003), arXiv: \eprint{astro-ph/0208251}.

\bibitem{2009ApJ...692.1060V}
A.~{Vikhlinin}, A.~V. {Kravtsov}, R.~A. {Burenin}, H.~{Ebeling}, W.~R. {Forman}, A.~{Hornstrup} et~al., \apj \textbf{692}, 1060 (2009), arXiv: \eprint{0812.2720}.

\bibitem{2017MNRAS.470..551Z}
Y.~{Zu}, R.~{Mandelbaum}, M.~{Simet}, E.~{Rozo}, and E.~S. {Rykoff}, \mnras \textbf{470}, 551 (2017), arXiv: \eprint{1611.00366}.

\bibitem{2018A&A...620A...1M}
F.~{Marulli}, A.~{Veropalumbo}, M.~{Sereno}, L.~{Moscardini}, F.~{Pacaud}, M.~{Pierre} et~al., \aap \textbf{620}, A1 (2018), arXiv: \eprint{1807.04760}.

\bibitem{2019MNRAS.482.1352M}
T.~{McClintock}, T.~N. {Varga}, D.~{Gruen}, E.~{Rozo}, E.~S. {Rykoff}, T.~{Shin} et~al., \mnras \textbf{482}, 1352 (2019), arXiv: \eprint{1805.00039}.

\bibitem{2019ApJ...878...55B}
S.~{Bocquet}, J.~P. {Dietrich}, T.~{Schrabback}, L.~E. {Bleem}, M.~{Klein}, S.~W. {Allen} et~al., \apj \textbf{878}, 55 (2019), arXiv: \eprint{1812.01679}.

\bibitem{2024A&A...682A.148F}
A.~{Fumagalli}, M.~{Costanzi}, A.~{Saro}, T.~{Castro}, and S.~{Borgani}, \aap \textbf{682}, A148 (2024), arXiv: \eprint{2310.09146}.

\bibitem{2011ARA&A..49..409A}
S.~W. {Allen}, A.~E. {Evrard}, and A.~B. {Mantz}, \araa \textbf{49}, 409 (2011), arXiv: \eprint{1103.4829}.

\bibitem{2012ARA&A..50..353K}
A.~V. {Kravtsov} and S.~{Borgani}, \araa \textbf{50}, 353 (2012), arXiv: \eprint{1205.5556}.

\bibitem{2013PhR...530...87W}
D.~H. {Weinberg}, M.~J. {Mortonson}, D.~J. {Eisenstein}, C.~{Hirata}, A.~G. {Riess}, and E.~{Rozo}, \physrep \textbf{530}, 87 (2013), arXiv: \eprint{1201.2434}.

\bibitem{2025arXiv250507697M}
H.~{Miyatake}, arXiv e-prints arXiv:2505.07697 (2025), arXiv: \eprint{2505.07697}.

\bibitem{2014ApJ...785..104R}
E.~S. {Rykoff}, E.~{Rozo}, M.~T. {Busha}, C.~E. {Cunha}, A.~{Finoguenov}, A.~{Evrard} et~al., \apj \textbf{785}, 104 (2014), arXiv: \eprint{1303.3562}.

\bibitem{2022PhRvD.105b3520A}
T.~M.~C. {Abbott}, M.~{Aguena}, A.~{Alarcon}, S.~{Allam}, O.~{Alves}, A.~{Amon} et~al., \prd \textbf{105}, 023520 (2022), arXiv: \eprint{2105.13549}.

\bibitem{2022A&A...659A..88L}
G.~F. {Lesci}, F.~{Marulli}, L.~{Moscardini}, M.~{Sereno}, A.~{Veropalumbo}, M.~{Maturi} et~al., \aap \textbf{659}, A88 (2022), arXiv: \eprint{2012.12273}.

\bibitem{2007ApJS..172..561B}
R.~A. {Burenin}, A.~{Vikhlinin}, A.~{Hornstrup}, H.~{Ebeling}, H.~{Quintana}, and A.~{Mescheryakov}, \apjs \textbf{172}, 561 (2007), arXiv: \eprint{astro-ph/0610739}.

\bibitem{2016A&A...592A...2P}
F.~{Pacaud}, N.~{Clerc}, P.~A. {Giles}, C.~{Adami}, T.~{Sadibekova}, M.~{Pierre} et~al., \aap \textbf{592}, A2 (2016), arXiv: \eprint{1512.04264}.

\bibitem{2022A&A...661A...7B}
Y.~E. {Bahar}, E.~{Bulbul}, N.~{Clerc}, V.~{Ghirardini}, A.~{Liu}, K.~{Nandra} et~al., \aap \textbf{661}, A7 (2022), arXiv: \eprint{2110.09534}.

\bibitem{2016A&A...594A..27P}
{Planck Collaboration}, P.~A.~R. {Ade}, N.~{Aghanim}, M.~{Arnaud}, M.~{Ashdown}, J.~{Aumont} et~al., \aap \textbf{594}, A27 (2016), arXiv: \eprint{1502.01598}.

\bibitem{2018ApJS..235...20H}
M.~{Hilton}, M.~{Hasselfield}, C.~{Sif{\'o}n}, N.~{Battaglia}, S.~{Aiola}, V.~{Bharadwaj} et~al., \apjs \textbf{235}, 20 (2018), arXiv: \eprint{1709.05600}.

\bibitem{2020AJ....159..110H}
N.~{Huang}, L.~E. {Bleem}, B.~{Stalder}, P.~A.~R. {Ade}, S.~W. {Allen}, A.~J. {Anderson} et~al., \aj \textbf{159}, 110 (2020), arXiv: \eprint{1907.09621}.

\bibitem{2015MNRAS.446.2205M}
A.~B. {Mantz}, A.~{von der Linden}, S.~W. {Allen}, D.~E. {Applegate}, P.~L. {Kelly}, R.~G. {Morris} et~al., \mnras \textbf{446}, 2205 (2015), arXiv: \eprint{1407.4516}.

\bibitem{2019MNRAS.489..401Z}
{\'I}.~{Zubeldia} and A.~{Challinor}, \mnras \textbf{489}, 401 (2019), arXiv: \eprint{1904.07887}.

\bibitem{2021PhRvD.103d3522C}
M.~{Costanzi}, A.~{Saro}, S.~{Bocquet}, T.~M.~C. {Abbott}, M.~{Aguena}, S.~{Allam} et~al., \prd \textbf{103}, 043522 (2021), arXiv: \eprint{2010.13800}.

\bibitem{2009arXiv0912.0201L}
{LSST Science Collaboration}, P.~A. {Abell}, J.~{Allison}, S.~F. {Anderson}, J.~R. {Andrew}, J.~R.~P. {Angel} et~al., arXiv e-prints arXiv:0912.0201 (2009), arXiv: \eprint{0912.0201}.

\bibitem{2011arXiv1110.3193L}
R.~{Laureijs}, J.~{Amiaux}, S.~{Arduini}, J.~L. {Augu{\`e}res}, J.~{Brinchmann}, R.~{Cole} et~al., arXiv e-prints arXiv:1110.3193 (2011), arXiv: \eprint{1110.3193}.

\bibitem{2024arXiv240513491E}
{Euclid Collaboration}, Y.~{Mellier}, {Abdurro'uf}, J.~A. {Acevedo Barroso}, A.~{Ach{\'u}carro}, J.~{Adamek} et~al., arXiv e-prints arXiv:2405.13491 (2024), arXiv: \eprint{2405.13491}.

\bibitem{2019BAAS...51c.341D}
O.~{Dore}, C.~{Hirata}, Y.~{Wang}, D.~{Weinberg}, T.~{Eifler}, R.~J. {Foley} et~al., \baas \textbf{51}, 341 (2019), arXiv: \eprint{1904.01174}.

\bibitem{2019ApJ...883..203G}
Y.~{Gong}, X.~{Liu}, Y.~{Cao}, X.~{Chen}, Z.~{Fan}, R.~{Li} et~al., \apj \textbf{883}, 203 (2019), arXiv: \eprint{1901.04634}.

\bibitem{2023MNRAS.519.1132M}
H.~{Miao}, Y.~{Gong}, X.~{Chen}, Z.~{Huang}, X.-D. {Li}, and H.~{Zhan}, \mnras \textbf{519}, 1132 (2023), arXiv: \eprint{2206.09822}.

\bibitem{2023RAA....23d5011Z}
Y.~{Zhang}, M.~{Chen}, Z.~{Wen}, and W.~{Fang}, Research in Astronomy and Astrophysics \textbf{23}, 045011 (2023), arXiv: \eprint{2302.05010}.

\bibitem{2025SCPMA..6880402G}
Y.~{Gong}, H.~{Miao}, X.~{Zhou}, Q.~{Xiong}, Y.~{Song}, Y.~{Jiang} et~al., Science China Physics, Mechanics, and Astronomy \textbf{68}, 280402 (2025), arXiv: \eprint{2501.15023}.

\bibitem{2025arXiv250704618C}
{CSST Collaboration}, Y.~{Gong}, H.~{Miao}, H.~{Zhan}, Z.-Y. {Li}, J.~{Shangguan} et~al., arXiv e-prints arXiv:2507.04618 (2025), arXiv: \eprint{2507.04618}.

\bibitem{2024AstTI...1...16J}
{JUST Team}, C.~{Liu}, Y.~{Zu}, F.~{Feng}, Z.~{Li}, Y.~{Yu} et~al., Astronomical Techniques and Instruments \textbf{1}, 16 (2024), arXiv: \eprint{2402.14312}.

\bibitem{1974ApJ...187..425P}
W.~H. {Press} and P.~{Schechter}, \apj \textbf{187}, 425 (1974).

\bibitem{1991ApJ...379..440B}
J.~R. {Bond}, S.~{Cole}, G.~{Efstathiou}, and N.~{Kaiser}, \apj \textbf{379}, 440 (1991).

\bibitem{2001MNRAS.321..372J}
A.~{Jenkins}, C.~S. {Frenk}, S.~D.~M. {White}, J.~M. {Colberg}, S.~{Cole}, A.~E. {Evrard} et~al., \mnras \textbf{321}, 372 (2001), arXiv: \eprint{astro-ph/0005260}.

\bibitem{2003MNRAS.346..565R}
D.~{Reed}, J.~{Gardner}, T.~{Quinn}, J.~{Stadel}, M.~{Fardal}, G.~{Lake} et~al., \mnras \textbf{346}, 565 (2003), arXiv: \eprint{astro-ph/0301270}.

\bibitem{2010MNRAS.403.1353C}
M.~{Crocce}, P.~{Fosalba}, F.~J. {Castander}, and E.~{Gazta{\~n}aga}, \mnras \textbf{403}, 1353 (2010), arXiv: \eprint{0907.0019}.

\bibitem{2012MNRAS.426.2046A}
R.~E. {Angulo}, V.~{Springel}, S.~D.~M. {White}, A.~{Jenkins}, C.~M. {Baugh}, and C.~S. {Frenk}, \mnras \textbf{426}, 2046 (2012), arXiv: \eprint{1203.3216}.

\bibitem{2013MNRAS.433.1230W}
W.~A. {Watson}, I.~T. {Iliev}, A.~{D'Aloisio}, A.~{Knebe}, P.~R. {Shapiro}, and G.~{Yepes}, \mnras \textbf{433}, 1230 (2013), arXiv: \eprint{1212.0095}.

\bibitem{2016MNRAS.456.2361B}
S.~{Bocquet}, A.~{Saro}, K.~{Dolag}, and J.~J. {Mohr}, \mnras \textbf{456}, 2361 (2016), arXiv: \eprint{1502.07357}.

\bibitem{2008ApJ...688..709T}
J.~{Tinker}, A.~V. {Kravtsov}, A.~{Klypin}, K.~{Abazajian}, M.~{Warren}, G.~{Yepes} et~al., \apj \textbf{688}, 709 (2008), arXiv: \eprint{0803.2706}.

\bibitem{2016MNRAS.456.2486D}
G.~{Despali}, C.~{Giocoli}, R.~E. {Angulo}, G.~{Tormen}, R.~K. {Sheth}, G.~{Baso} et~al., \mnras \textbf{456}, 2486 (2016), arXiv: \eprint{1507.05627}.

\bibitem{2020ApJ...903...87D}
B.~{Diemer}, \apj \textbf{903}, 87 (2020), arXiv: \eprint{2007.10346}.

\bibitem{2022MNRAS.509.6077O}
L.~{Ondaro-Mallea}, R.~E. {Angulo}, M.~{Zennaro}, S.~{Contreras}, and G.~{Aric{\`o}}, \mnras \textbf{509}, 6077 (2022), arXiv: \eprint{2102.08958}.

\bibitem{2020A&A...643A..20S}
L.~{Salvati}, M.~{Douspis}, and N.~{Aghanim}, \aap \textbf{643}, A20 (2020), arXiv: \eprint{2005.10204}.

\bibitem{2021A&A...649A..47A}
E.~{Artis}, J.-B. {Melin}, J.~G. {Bartlett}, and C.~{Murray}, \aap \textbf{649}, A47 (2021), arXiv: \eprint{2101.02501}.

\bibitem{2023A&A...671A.100E}
{Euclid Collaboration}, T.~{Castro}, A.~{Fumagalli}, R.~E. {Angulo}, S.~{Bocquet}, S.~{Borgani} et~al., \aap \textbf{671}, A100 (2023), arXiv: \eprint{2208.02174}.

\bibitem{2011ApJ...732..122B}
S.~{Bhattacharya}, K.~{Heitmann}, M.~{White}, Z.~{Luki{\'c}}, C.~{Wagner}, and S.~{Habib}, \apj \textbf{732}, 122 (2011), arXiv: \eprint{1005.2239}.

\bibitem{2025JCAP...02..021A}
A.~G. {Adame}, J.~{Aguilar}, S.~{Ahlen}, S.~{Alam}, D.~M. {Alexander}, M.~{Alvarez} et~al., \jcap \textbf{2025}, 021 (2025), arXiv: \eprint{2404.03002}.

\bibitem{2025arXiv250314738D}
{DESI Collaboration}, M.~{Abdul-Karim}, J.~{Aguilar}, S.~{Ahlen}, S.~{Alam}, L.~{Allen} et~al., arXiv e-prints arXiv:2503.14738 (2025), arXiv: \eprint{2503.14738}.

\bibitem{2022LRCA....8....1A}
R.~E. {Angulo} and O.~{Hahn}, Living Reviews in Computational Astrophysics \textbf{8}, 1 (2022), arXiv: \eprint{2112.05165}.

\bibitem{2023RPPh...86g6901M}
K.~{Moriwaki}, T.~{Nishimichi}, and N.~{Yoshida}, Reports on Progress in Physics \textbf{86}, 076901 (2023), arXiv: \eprint{2303.15794}.

\bibitem{2019ApJ...872...53M}
T.~{McClintock}, E.~{Rozo}, M.~R. {Becker}, J.~{DeRose}, Y.-Y. {Mao}, S.~{McLaughlin} et~al., \apj \textbf{872}, 53 (2019), arXiv: \eprint{1804.05866}.

\bibitem{2019ApJ...884...29N}
T.~{Nishimichi}, M.~{Takada}, R.~{Takahashi}, K.~{Osato}, M.~{Shirasaki}, T.~{Oogi} et~al., \apj \textbf{884}, 29 (2019), arXiv: \eprint{1811.09504}.

\bibitem{2020ApJ...901....5B}
S.~{Bocquet}, K.~{Heitmann}, S.~{Habib}, E.~{Lawrence}, T.~{Uram}, N.~{Frontiere} et~al., \apj \textbf{901}, 5 (2020), arXiv: \eprint{2003.12116}.

\bibitem{2025JCAP...03..056S}
D.~{Shen}, N.~{Kokron}, J.~{DeRose}, J.~{Tinker}, R.~H. {Wechsler}, A.~{Banerjee} et~al., \jcap \textbf{2025}, 056 (2025), arXiv: \eprint{2410.00913}.

\bibitem{2010ApJ...715..104H}
K.~{Heitmann}, M.~{White}, C.~{Wagner}, S.~{Habib}, and D.~{Higdon}, \apj \textbf{715}, 104 (2010), arXiv: \eprint{0812.1052}.

\bibitem{2014ApJ...780..111H}
K.~{Heitmann}, E.~{Lawrence}, J.~{Kwan}, S.~{Habib}, and D.~{Higdon}, \apj \textbf{780}, 111 (2014), arXiv: \eprint{1304.7849}.

\bibitem{2019MNRAS.484.5509E}
{Euclid Collaboration}, M.~{Knabenhans}, J.~{Stadel}, S.~{Marelli}, D.~{Potter}, R.~{Teyssier} et~al., \mnras \textbf{484}, 5509 (2019), arXiv: \eprint{1809.04695}.

\bibitem{2021MNRAS.505.2840E}
{Euclid Collaboration}, M.~{Knabenhans}, J.~{Stadel}, D.~{Potter}, J.~{Dakin}, S.~{Hannestad} et~al., \mnras \textbf{505}, 2840 (2021), arXiv: \eprint{2010.11288}.

\bibitem{2023MNRAS.520.3443M}
K.~R. {Moran}, K.~{Heitmann}, E.~{Lawrence}, S.~{Habib}, D.~{Bingham}, A.~{Upadhye} et~al., \mnras \textbf{520}, 3443 (2023), arXiv: \eprint{2207.12345}.

\bibitem{Chen2025}
Z.~{Chen}, Y.~{Yu}, J.~{Han}, and Y.~{Jing}, Science China Physics, Mechanics, and Astronomy \textbf{68}, 289512 (2025), arXiv: \eprint{2502.11160}.

\bibitem{2019arXiv190713167M}
T.~{McClintock}, E.~{Rozo}, A.~{Banerjee}, M.~R. {Becker}, J.~{DeRose}, S.~{McLaughlin} et~al., arXiv e-prints arXiv:1907.13167 (2019), arXiv: \eprint{1907.13167}.

\bibitem{2015ApJ...810...35K}
J.~{Kwan}, K.~{Heitmann}, S.~{Habib}, N.~{Padmanabhan}, E.~{Lawrence}, H.~{Finkel} et~al., \apj \textbf{810}, 35 (2015), arXiv: \eprint{1311.6444}.

\bibitem{2019ApJ...874...95Z}
Z.~{Zhai}, J.~L. {Tinker}, M.~R. {Becker}, J.~{DeRose}, Y.-Y. {Mao}, T.~{McClintock} et~al., \apj \textbf{874}, 95 (2019), arXiv: \eprint{1804.05867}.

\bibitem{2020MNRAS.492.2872W}
B.~D. {Wibking}, D.~H. {Weinberg}, A.~N. {Salcedo}, H.-Y. {Wu}, S.~{Singh}, S.~{Rodr{\'\i}guez-Torres} et~al., \mnras \textbf{492}, 2872 (2020), arXiv: \eprint{1907.06293}.

\bibitem{2022MNRAS.515..871Y}
S.~{Yuan}, L.~H. {Garrison}, D.~J. {Eisenstein}, and R.~H. {Wechsler}, \mnras \textbf{515}, 871 (2022), arXiv: \eprint{2203.11963}.

\bibitem{2023ApJ...948...99Z}
Z.~{Zhai}, J.~L. {Tinker}, A.~{Banerjee}, J.~{DeRose}, H.~{Guo}, Y.-Y. {Mao} et~al., \apj \textbf{948}, 99 (2023), arXiv: \eprint{2203.08999}.

\bibitem{2024ApJ...961..208S}
K.~{Storey-Fisher}, J.~L. {Tinker}, Z.~{Zhai}, J.~{DeRose}, R.~H. {Wechsler}, and A.~{Banerjee}, \apj \textbf{961}, 208 (2024), arXiv: \eprint{2210.03203}.

\bibitem{2021JCAP...09..020H}
B.~{Hadzhiyska}, C.~{Garc{\'\i}a-Garc{\'\i}a}, D.~{Alonso}, A.~{Nicola}, and A.~{Slosar}, \jcap \textbf{2021}, 020 (2021), arXiv: \eprint{2103.09820}.

\bibitem{2023MNRAS.524.2407Z}
M.~{Zennaro}, R.~E. {Angulo}, M.~{Pellejero-Ib{\'a}{\~n}ez}, J.~{St{\"u}cker}, S.~{Contreras}, and G.~{Aric{\`o}}, \mnras \textbf{524}, 2407 (2023), arXiv: \eprint{2101.12187}.

\bibitem{2023MNRAS.520.3725P}
M.~{Pellejero Iba{\~n}ez}, R.~E. {Angulo}, M.~{Zennaro}, J.~{St{\"u}cker}, S.~{Contreras}, G.~{Aric{\`o}} et~al., \mnras \textbf{520}, 3725 (2023), arXiv: \eprint{2207.06437}.

\bibitem{2023JCAP...07..054D}
J.~{DeRose}, N.~{Kokron}, A.~{Banerjee}, S.-F. {Chen}, M.~{White}, R.~{Wechsler} et~al., \jcap \textbf{2023}, 054 (2023), arXiv: \eprint{2303.09762}.

\bibitem{2025arXiv250604671Z}
S.~{Zhou}, Z.~{Chen}, and Y.~{Yu}, arXiv e-prints arXiv:2506.04671 (2025), arXiv: \eprint{2506.04671}.

\bibitem{2018JCAP...03..049L}
J.~{Liu}, S.~{Bird}, J.~M. {Zorrilla Matilla}, J.~C. {Hill}, Z.~{Haiman}, M.~S. {Madhavacheril} et~al., \jcap \textbf{2018}, 049 (2018), arXiv: \eprint{1711.10524}.

\bibitem{2019PhRvD..99h3508L}
J.~{Liu} and M.~S. {Madhavacheril}, \prd \textbf{99}, 083508 (2019), arXiv: \eprint{1809.10747}.

\bibitem{2019PhRvD..99f3527L}
Z.~{Li}, J.~{Liu}, J.~M. {Zorrilla Matilla}, and W.~R. {Coulton}, \prd \textbf{99}, 063527 (2019), arXiv: \eprint{1810.01781}.

\bibitem{2019JCAP...05..043C}
W.~R. {Coulton}, J.~{Liu}, M.~S. {Madhavacheril}, V.~{B{\"o}hm}, and D.~N. {Spergel}, \jcap \textbf{2019}, 043 (2019), arXiv: \eprint{1810.02374}.

\bibitem{2025PhRvD.111d1302W}
Y.~{Wang} and P.~{He}, \prd \textbf{111}, L041302 (2025), arXiv: \eprint{2408.13876}.

\bibitem{1998ApJ...495...80B}
G.~L. {Bryan} and M.~L. {Norman}, \apj \textbf{495}, 80 (1998), arXiv: \eprint{astro-ph/9710107}.

\bibitem{2014JCAP...02..049C}
E.~{Castorina}, E.~{Sefusatti}, R.~K. {Sheth}, F.~{Villaescusa-Navarro}, and M.~{Viel}, \jcap \textbf{2014}, 049 (2014), arXiv: \eprint{1311.1212}.

\bibitem{1996ApJ...469..480K}
T.~{Kitayama} and Y.~{Suto}, \apj \textbf{469}, 480 (1996), arXiv: \eprint{astro-ph/9604141}.

\bibitem{2011MNRAS.410.1911C}
J.~{Courtin}, Y.~{Rasera}, J.~M. {Alimi}, P.~S. {Corasaniti}, V.~{Boucher}, and A.~{F{\"u}zfa}, \mnras \textbf{410}, 1911 (2011), arXiv: \eprint{1001.3425}.

\bibitem{2024arXiv241118722L}
Y.~{Li} and R.~E. {Smith}, arXiv e-prints arXiv:2411.18722 (2024), arXiv: \eprint{2411.18722}.

\bibitem{2024MNRAS.535.3500Y}
S.~{Yue}, L.~{Feng}, W.~{Ju}, J.~{Pan}, Z.~{Huang}, F.~{Fang} et~al., \mnras \textbf{535}, 3500 (2024), arXiv: \eprint{2408.16398}.

\bibitem{2020ApJS..250....2V}
F.~{Villaescusa-Navarro}, C.~{Hahn}, E.~{Massara}, A.~{Banerjee}, A.~M. {Delgado}, D.~K. {Ramanah} et~al., \apjs \textbf{250}, 2 (2020), arXiv: \eprint{1909.05273}.

\bibitem{Han2025}
J.~{Han}, M.~{Li}, W.~{Jiang}, Z.~{Chen}, H.~{Wang}, C.~{Wei} et~al., arXiv e-prints arXiv:2503.21368 (2025), arXiv: \eprint{2503.21368}.

\bibitem{2001IJMPD..10..213C}
M.~{Chevallier} and D.~{Polarski}, International Journal of Modern Physics D \textbf{10}, 213 (2001), arXiv: \eprint{gr-qc/0009008}.

\bibitem{2003PhRvL..90i1301L}
E.~V. {Linder}, \prl \textbf{90}, 091301 (2003), arXiv: \eprint{astro-ph/0208512}.

\bibitem{2020JCAP...09..018P}
C.~{Partmann}, C.~{Fidler}, C.~{Rampf}, and O.~{Hahn}, \jcap \textbf{2020}, 018 (2020), arXiv: \eprint{2003.07387}.

\bibitem{2022JCAP...09..068H}
P.~{Heuschling}, C.~{Partmann}, and C.~{Fidler}, \jcap \textbf{2022}, 068 (2022), arXiv: \eprint{2201.13186}.

\bibitem{sobol1967distribution}
I.~Sobol, Vychisl. Mat. i Mater. Phys \textbf{7}, 784 (1967).

\bibitem{mckayComparisonThreeMethods1979}
M.~D. McKay, R.~J. Beckman, and W.~J. Conover, Technometrics \textbf{21}, 239 (1979).

\bibitem{tangOrthogonalArrayBasedLatin1993}
B.~Tang, Journal of the American Statistical Association \textbf{88}, 1392 (1993).

\bibitem{2016MNRAS.462L...1A}
R.~E. {Angulo} and A.~{Pontzen}, \mnras \textbf{462}, L1 (2016), arXiv: \eprint{1603.05253}.

\bibitem{2013ApJ...762..109B}
P.~S. {Behroozi}, R.~H. {Wechsler}, and H.-Y. {Wu}, \apj \textbf{762}, 109 (2013), arXiv: \eprint{1110.4372}.

\bibitem{2001MNRAS.328..726S}
V.~{Springel}, S.~D.~M. {White}, G.~{Tormen}, and G.~{Kauffmann}, \mnras \textbf{328}, 726 (2001), arXiv: \eprint{astro-ph/0012055}.

\bibitem{2024A&A...691A.323S}
I.~{S{\'a}ez-Casares}, Y.~{Rasera}, T.~R.~G. {Richardson}, and P.~S. {Corasaniti}, \aap \textbf{691}, A323 (2024), arXiv: \eprint{2410.05226}.

\bibitem{2020SciPy-NMeth}
P.~Virtanen, R.~Gommers, T.~E. Oliphant, M.~Haberland, T.~Reddy, D.~Cournapeau et~al., Nature Methods \textbf{17}, 261 (2020).

\bibitem{C25}
T.~{Castro}, S.~{Borgani}, and J.~{Dakin}, arXiv e-prints arXiv:2504.07608 (2025), arXiv: \eprint{2504.07608}.

\bibitem{scikit-learn}
F.~Pedregosa, G.~Varoquaux, A.~Gramfort, V.~Michel, B.~Thirion, O.~Grisel et~al., Journal of Machine Learning Research \textbf{12}, 2825 (2011).

\bibitem{1984ApJ...281....1F}
J.~A. {Fillmore} and P.~{Goldreich}, \apj \textbf{281}, 1 (1984).

\bibitem{1985ApJS...58...39B}
E.~{Bertschinger}, \apjs \textbf{58}, 39 (1985).

\bibitem{2014JCAP...11..019A}
S.~{Adhikari}, N.~{Dalal}, and R.~T. {Chamberlain}, \jcap \textbf{2014}, 019 (2014), arXiv: \eprint{1409.4482}.

\bibitem{2014ApJ...789....1D}
B.~{Diemer} and A.~V. {Kravtsov}, \apj \textbf{789}, 1 (2014), arXiv: \eprint{1401.1216}.

\bibitem{2021MNRAS.503.4250F}
M.~{Fong} and J.~{Han}, \mnras \textbf{503}, 4250 (2021), arXiv: \eprint{2008.03477}.

\bibitem{2022MNRAS.513.4754F}
M.~{Fong}, J.~{Han}, J.~{Zhang}, X.~{Yang}, H.~{Gao}, J.~{Wang} et~al., \mnras \textbf{513}, 4754 (2022), arXiv: \eprint{2205.01816}.

\bibitem{2023ApJ...953...37G}
H.~{Gao}, J.~{Han}, M.~{Fong}, Y.~P. {Jing}, and Z.~{Li}, \apj \textbf{953}, 37 (2023), arXiv: \eprint{2303.10887}.

\bibitem{2023MNRAS.525.2489Z}
Y.~{Zhou} and J.~{Han}, \mnras \textbf{525}, 2489 (2023), arXiv: \eprint{2303.10886}.

\bibitem{2025ApJ...979...55Z}
Y.~{Zhou} and J.~{Han}, \apj \textbf{979}, 55 (2025), arXiv: \eprint{2407.08381}.

\bibitem{2019MNRAS.488.4779C}
M.~{Costanzi}, E.~{Rozo}, M.~{Simet}, Y.~{Zhang}, A.~E. {Evrard}, A.~{Mantz} et~al., \mnras \textbf{488}, 4779 (2019), arXiv: \eprint{1810.09456}.

\bibitem{2020Natur.585..357H}
C.~R. {Harris}, K.~J. {Millman}, S.~J. {van der Walt}, R.~{Gommers}, P.~{Virtanen}, D.~{Cournapeau} et~al., \nat \textbf{585}, 357 (2020), arXiv: \eprint{2006.10256}.

\end{thebibliography}



\end{multicols}
\end{document}